\documentclass[11pt]{article}
\RequirePackage{fix-cm}
\usepackage[utf8]{inputenc}
\usepackage[margin=1.5in]{geometry}
\usepackage[bottom, flushmargin]{footmisc}
\usepackage{titlesec}

\titlespacing\section{0pt}{0.21in}{0.03in}
\titlespacing\subsection{0pt}{0.21in}{0.01in}
\titlespacing\subsubsection{0pt}{0.21in}{0in}

\usepackage[superscript]{cite}
\usepackage{enumitem}

\usepackage{microtype}

\PassOptionsToPackage{hyphens}{url} 
\usepackage[hyperfootnotes=false]{hyperref} 
\usepackage{comment}
\usepackage{graphicx,nicefrac}
\usepackage{caption}
\graphicspath{ {dataviz/} }
\usepackage{amsmath}
\usepackage{amssymb}
\usepackage{bm}
\usepackage{float}
\usepackage{mathtools}

\usepackage{color}
\usepackage{xcolor, colortbl}

\hypersetup{
	colorlinks,
	linkcolor=black, 
	urlcolor=cyan,  
	citecolor= red, 
	breaklinks=true
}

\usepackage{multirow}
\usepackage{multicol}
\usepackage{array}
\usepackage{booktabs}
\newcolumntype{L}[1]{>{\raggedright\let\newline\\\arraybackslash\hspace{0pt}}m{#1}}
\newcolumntype{C}[1]{>{\centering\let\newline\\\arraybackslash\hspace{0pt}}m{#1}}
\newcolumntype{R}[1]{>{\raggedleft\let\newline\\\arraybackslash\hspace{0pt}}m{#1}}

\usepackage{marvosym}

\begin{document}
\pagenumbering{gobble}

\begin{center}
	\begin{minipage}{0.9\linewidth}
		\begin{center}
			\LARGE
			Who Moved? Ecological Estimates of \\Turnout \& Coalition Support by Ethnicity and Age\\in Johor, Malaysia, 2022--2026\\[0.21in]
			\Large
			Thevesh Thevananthan\footnotemark$^{\text{\Letter}}$ and Ong Kian Ming\footnotemark\\[0.15in]
		\end{center}
		\large
		At the July 2026 Johor state election, Barisan Nasional (BN) swept the state and nearly doubled its vote relative to 2022. We exploit geographic variation in the ethnic composition of polling stations and administration-induced variation in the age composition of polling streams to estimate turnout and coalition support by ethnicity and age for the three elections involving Johor since 2022, with five key findings. (1) Turnout is a first-order driver of outcomes, moving by up to 26 points within an ethnic group across elections and following a life-cycle profile. (2) BN's surge was a near one-for-one Perikatan Nasional (PN) collapse among Malay voters, robust to PN's stand-down in 23 of 56 seats. (3) The largest reversal towards BN was among young Malays, who had previously carried the `green wave' of 2022. (4) Two-thirds of Pakatan Harapan's (PH) decline among the Chinese electorate came through lower Chinese turnout; any crossover to BN was comparatively small. (5) The apparent youth gradient in BN's 2026 support is largely ethnic composition rather than within-group behaviour: by 2026, Malay support for every coalition was close to flat in age. More broadly, we provide what we believe are the first open and fully reproducible ecological estimates of voting behaviour for any Malaysian election, accompanied by estimand-specific validation errors measured against held-out behaviour.
	\end{minipage}
\end{center}

\footnotetext[1]{Universiti Malaya, W.P. Kuala Lumpur, Malaysia\\\Letter\space \href{mailto:academia@thevesh.com}{academia@thevesh.com}}
\footnotetext[2]{Taylor's University, Selangor, Malaysia}

\newpage
\pagenumbering{arabic}

\section{Introduction}
\label{sec:intro}

On 11 July 2026, Barisan Nasional (BN) won the 16th Johor state election (SE-16) in a landslide: 48 of 56 seats on a 59.7\% vote share, against 8 seats and 32.6\% for Pakatan Harapan (PH), its federal coalition partner. Perikatan Nasional (PN) won nothing, its 5.4\% statewide share reflecting a decision to contest only 33 of 56 seats (Section~\ref{sec:context}). BN's win was remarkable not just for its magnitude, but also its trajectory---it near-doubled its share of Johor's vote relative to the 15th general election (GE-15) in November 2022, which was also the worst federal result for BN in its history.

What the aggregate returns cannot say is \emph{who moved} and thus generated this result. Did Malay voters who defected to PN in 2022 return to UMNO, BN's anchor party? Did Chinese voters, PH's most loyal voter base, cross over to BN---or stay home? Did youth voters swing with the tide or against it? These are important questions, but they are also the kind commentators tend to answer with constituency-level correlations---exactly the inferences Robinson\cite{robinson1950} showed can be arbitrarily wrong.

We answer them with ecological inference (EI) at the finest level Malaysian data permit: the \textit{saluran}, or polling stream, of which Johor had 4,638--4,889 per election. For each saluran we observe two margins of a contingency table exactly---the demographic composition of its electorate and its full vote count---while the interior is unobserved. We estimate the interior for three elections run on identical administrative machinery (the March 2022 state election SE-15, GE-15, and SE-16), thus obtaining a comparable series of turnout and coalition support by ethnicity and age.

Our paper makes three contributions. First, the estimates themselves are, to the best of our knowledge, the first formal, open, and fully reproducible saluran-level estimates of voting behaviour by demographic group for Malaysia. Existing quantitative work operates at the constituency level \cite{pepinsky2015,dettmanpepinsky2024,chen2026}, relies on national surveys \cite{marzukisuffian2023}, or reaches the saluran level only through descriptive or proprietary analyses with no explicit estimator, no quantification of uncertainty, and no open release of data or code for replication\cite{zhanghutchinson2022,welsh2026b}. Our entire pipeline---data construction, estimation, validation, and presentation---is released and archived for replication.

Second, the findings surface subtleties underlying BN's comeback. The first is participation itself: turnout is a first-order driver of Johorean outcomes, moving by up to 26 points within an ethnic group across elections and following a life-cycle profile within each. Among the Malay electorate, we find that BN support jumped from 36.5\% at GE-15 to 64.1\% at SE-16, while PN support fell from 34.5\% to 6.8\%---a near one-for-one reversal, robust to PN's stand-down in 23 of 56 seats. This PN--BN reversal was steepest among young Malay voters; among Malays, PN won 52.7\% of 18--29 ballots at GE-15 (BN: 34.7\%), against just 7.6\% (BN: 86.8\%) at SE-16. In essence, BN reversed the youth-led `green wave'. That said, from a cross-sectional perspective, the apparent youth preference for BN in 2026 is largely composition: conditional on ethnicity, support for BN is generally flat in age---whereas the 2022 gradient was genuinely young Malays tilting to PN. We also find that the drop in Chinese support for PH (from 63.1\% of the Chinese electorate at GE-15 to 52.2\% at SE-16) is primarily attributable to demobilisation rather than crossover to BN; Chinese voter turnout dropped by 8 percentage points, accounting for two-thirds of PH's loss among the Chinese electorate.

Third, we leverage the presence of `demographically pure' salurans---created by geographic variation in ethnic composition and administrative sorting of voters into salurans based on age---to validate our estimates in a way that EI applications generally find infeasible. EI's reliability has been contested since the method existed, because its identifying assumption is untestable in the data\cite{freedman1998,cho1998}, and validations against known interiors are rare \cite{hudson2010,park2014,plesciadesio2018,pavia2022}. The aforementioned variation in our data allows us to conduct out-of-sample validation against these demographically pure salurans, thus giving us a rigorous way of quantifying the uncertainty inherent in our estimates. Rather than standard errors---which quantify precision rather than accuracy---we report the measured distance between prediction and observed behaviour, specific to each estimand. For groups with no demographically pure salurans (Indian and Other voters), no validation is possible; we report their point estimates without error bars and flag them as not citable.

The rest of our paper proceeds as follows. Section~\ref{sec:context} contextualises the three studied elections and establishes their comparability. Section~\ref{sec:data} describes the data. Section~\ref{sec:strategy} describes our empirical strategy. Section~\ref{sec:findings} presents the estimates. Section~\ref{sec:implications} draws out generalisable aspects of our methodology and findings. Section~\ref{sec:conclusion} concludes.

\section{Three elections in four years: Johor, 2022--2026}
\label{sec:context}

\begin{figure}[!htb]
	\centering
	\caption{Coalition vote share and turnout across the three elections}
	\includegraphics[width=\textwidth]{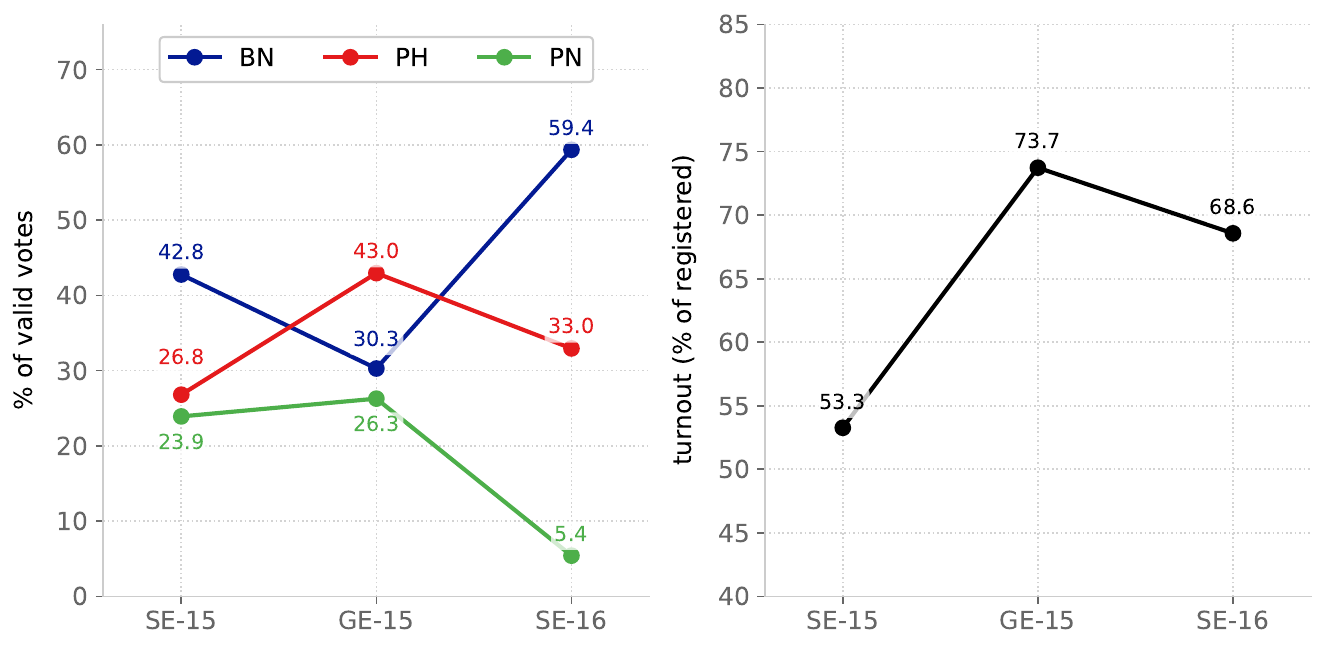}
	\label{fig:trajectory}
\end{figure}

Demography has underpinned Malaysian electoral competition since before independence \cite{ratnam1965,horowitz1985}: the major coalitions are alliances of ethnically based parties, and every election is read as a census of communal alignments---which is why this paper's questions are posed in ethnic and generational terms. Johor, the birthplace of UMNO and for six decades the party's safest large state \cite{hutchinson2018}, delivered two shocks in quick succession: BN lost the state (and the federal government) at the 14th general election of 2018 amid the 1MDB scandal \cite{funston2018,hutchinsonlee2019}, and the PH state government that followed fell in the `Sheraton Move' of early 2020, leaving a slim BN-led administration.

That administration called the 15th state election of 12 March 2022 (SE-15) early \cite{hutchinsonzhang2022a}. It was the first election under two franchise reforms---the `Undi18' lowering of the voting age to 18, and automatic registration---which together expanded Johor's roll by roughly 40\% \cite{chai2022}. It was also the last election of Malaysia's pandemic era: travel prior to 1 April 2022 was still restricted, a constraint that fell hardest on Johor's large cross-border workforce in Singapore \cite{malaymail2022border}. BN won 40 of 56 seats on 43\% of the vote and turnout of 55\%, the lowest of the three elections. Eight months later, at GE-15 (19 November 2022), the same electorate met the `green wave'---the national surge of the Malay-Islamist PN that produced a hung parliament \cite{chin2023,washida2023}. The deadlock was resolved by a `unity government' under PH leader Anwar Ibrahim, which UMNO/BN joined, leaving PN as the sole major opposition bloc.

SE-16, on 11 July 2026, was thus the first Johor-wide test of a party system reshaped twice over: BN and PH governing together federally while contesting each other for the state, and PN attempting to convert its 2022 momentum into state power. The intervening years treated the blocs very differently: the BN state government rode Johor's economic upswing and dissolved the assembly roughly a year early \cite{hutchinson2026}, while PN came apart in the months before the poll \cite{fulcrum2026}; two former PH ministers ran a minor party, BERSAMA, in 15 seats \cite{fmt2026}. PN fielded candidates in only 33 of the 56 seats---a deliberate choice, made clear by observing that the 33 it contested are 63.7\% Malay and 27.6\% Chinese, while the 23 it vacated (explicitly endorsing BN in the process) are 43.8\% Malay and 45.6\% Chinese.

The result was a BN landslide (Figure~\ref{fig:trajectory}): 48 of 56 seats on 59.7\%, its best Johor result since 2008. PH held 8 seats, its worst Johor result since 2013. PN lost every seat it contested, its share in those 33 seats falling from 31.6\% to 10.3\%. Turnout was 69.6\%: fourteen points above SE-15, six below GE-15.

Two features of this sequence shape everything that follows. First, the three elections were run on identical administrative machinery---the same polling districts, with no changes to their codes, the same age-based sorting rules, and rolls produced by the same system following the franchise reforms. The estimates are therefore built on a complete panel of directly comparable reporting units, without requiring geographic crosswalks or any other harmonisation assumption. This is comparability of measurement, not of politics: election type, stakes and the choice set still differed across the three contests, so a difference between them describes what happened rather than realignment net of type. Second, the sharp differences in turnout across the three elections (53.3\%, 73.7\%, 68.6\%) underscore the necessity of estimating turnout and vote choice jointly, and of reporting support both as a share of the electorate and as a share of ballots cast: when participation moves by twenty points, `Who moved?' needs to be put in context of `Who showed up?'.

\section{Data}
\label{sec:data}

To operationalise voting, the Election Commission (EC) assigns every Malaysian voter to a polling district (\emph{daerah mengundi}), and within it to a numbered polling stream (\emph{saluran}) within a polling station (\emph{pusat mengundi})---this is usually a single classroom or hall with its own ballot box, its own voter list (the \emph{Daftar Pemilih Pilihan Raya}), and its own published vote count (via Form 14).

For each election, we use data provided by Malaysian Election Corpus \cite{meco1} and join two datasets at saluran level. The first is the electoral roll, which records each voter's ethnicity and birth year, thus allowing us to provide exact counts in four ethnic groups (Malay, Chinese, Indian, Other) and six age bands (18--29, 30--39, 40--49, 50--59, 60--69, 70+). The second is the saluran-level results, which enumerate the number of ballots issued, votes by candidate, as well as rejected and unreturned ballots. It is worth noting that the saluran-level results (Form 14) published by the EC do not contain voter counts; it is only possible to derive this by joining against the voter roll.

\begin{figure}[!htb]
	\centering
	\caption{Why only Malay and Chinese rates are identified}
	\includegraphics[width=\textwidth]{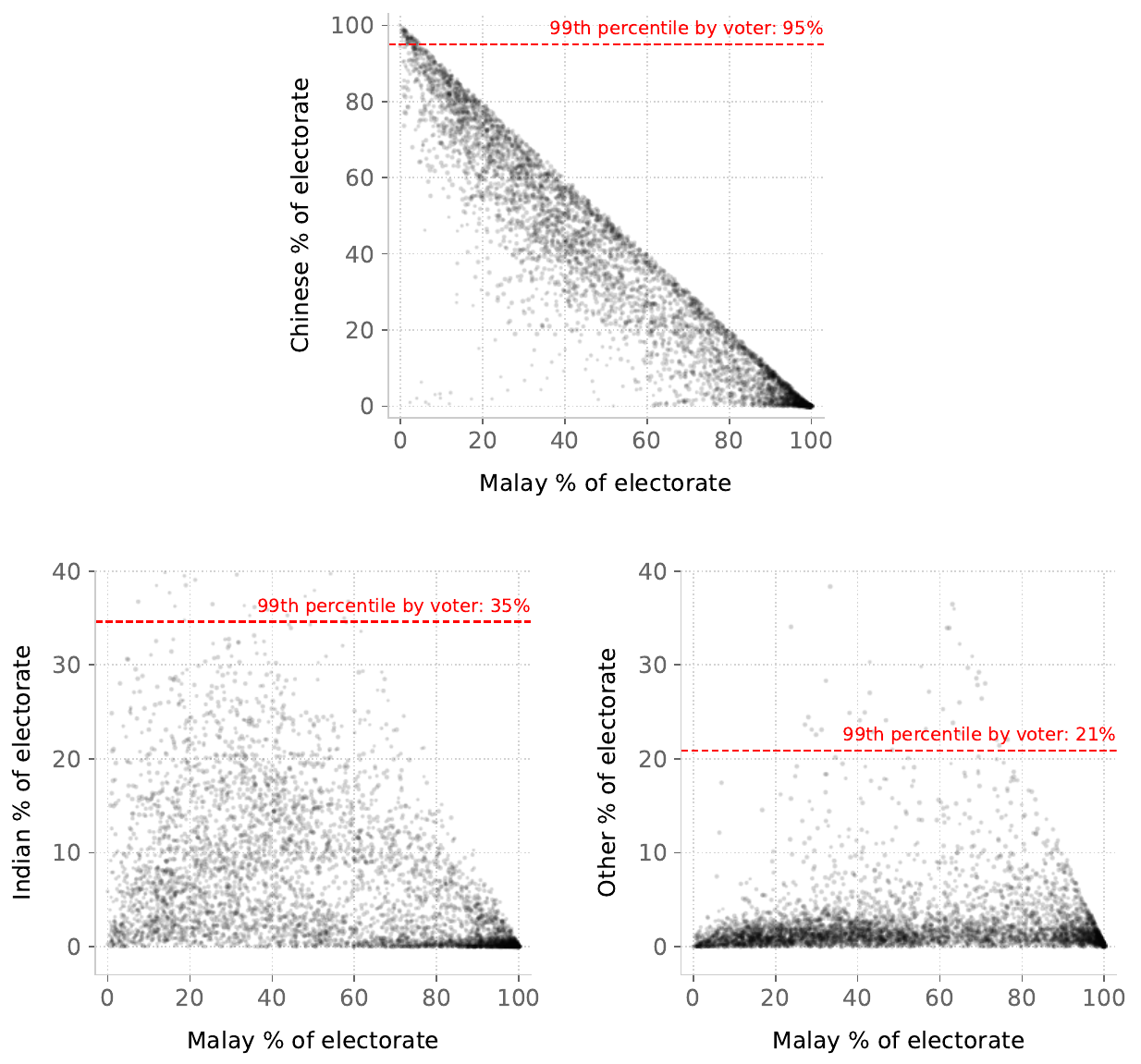}
	\label{fig:collinearity}
\end{figure}

Two independent sources of compositional variation across salurans do the identifying work in everything that follows. The first is geographic. Malay and Chinese shares trade off almost one-for-one across salurans---their correlation is close to $-1$---so the data are rich in salurans dominated by either group, and identification for their rates is strong. No saluran, by contrast, is remotely dominated by Indian or Other voters: the voter-weighted 99th percentile of Indian share is about 35\%, of Other share about 21\% (Figure~\ref{fig:collinearity}), so any estimate of their behaviour is an extrapolation far beyond the support of the data. This is a feature of Johor's residential geography, not a defect an estimator can repair.

\begin{figure}[!htb]
	\centering
	\caption{The Election Commission sorts voters into streams by age}
	\includegraphics[width=\textwidth]{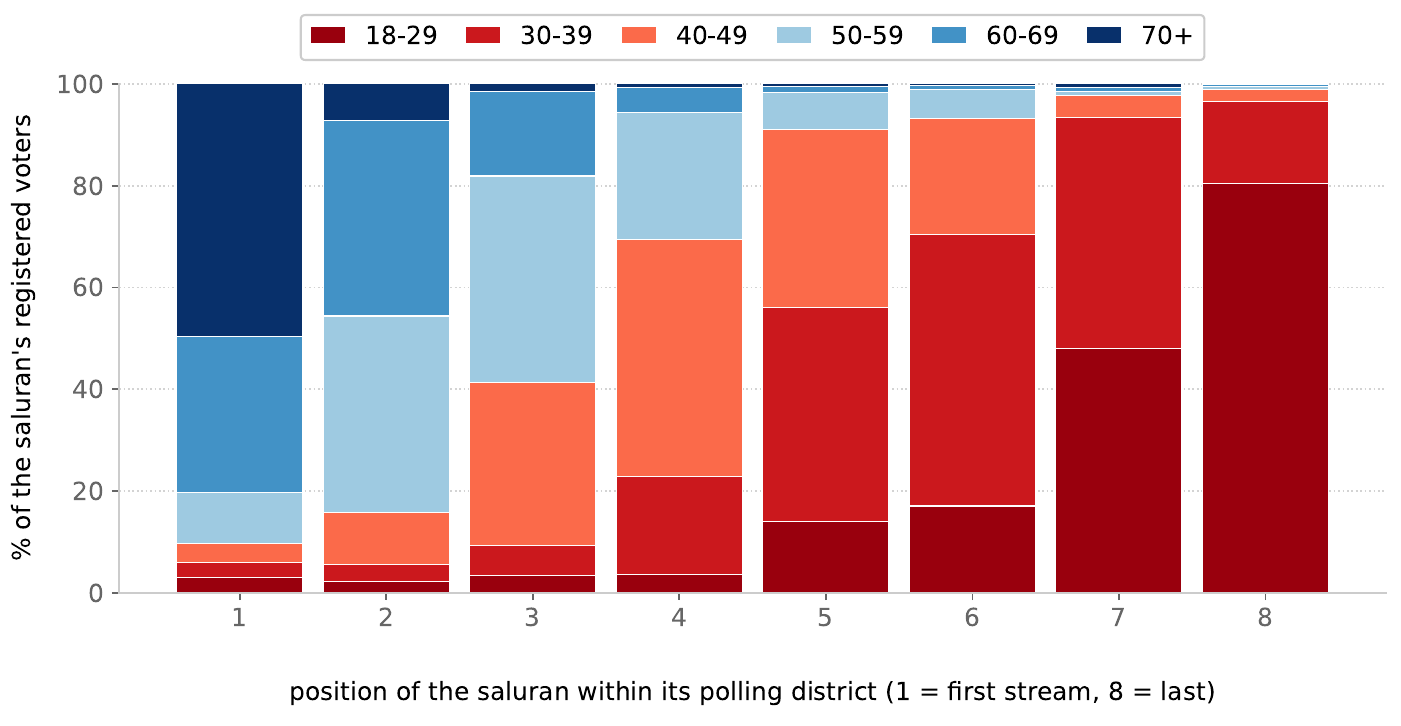}
	\label{fig:sorting}
\end{figure}

The second is administrative. The EC assigns voters to salurans within a polling district by age: saluran 1 receives the oldest voters, the last the youngest. Figure~\ref{fig:sorting} shows how sharply the rule sorts: in the first stream of a district, half the registered voters are 70 or over; in the last, four fifths are under 30. Each source of variation also yields near-pure salurans---ethnically pure ones from residential geography, age-pure ones from the sorting rule---in which a group's behaviour is very nearly observed, and against which Section~\ref{sec:strategy-groundtruth} validates the estimates.

\begin{figure}[!htb]
	\centering
	\caption{Ethnic composition of each age band, SE-16}
	\includegraphics[width=\textwidth]{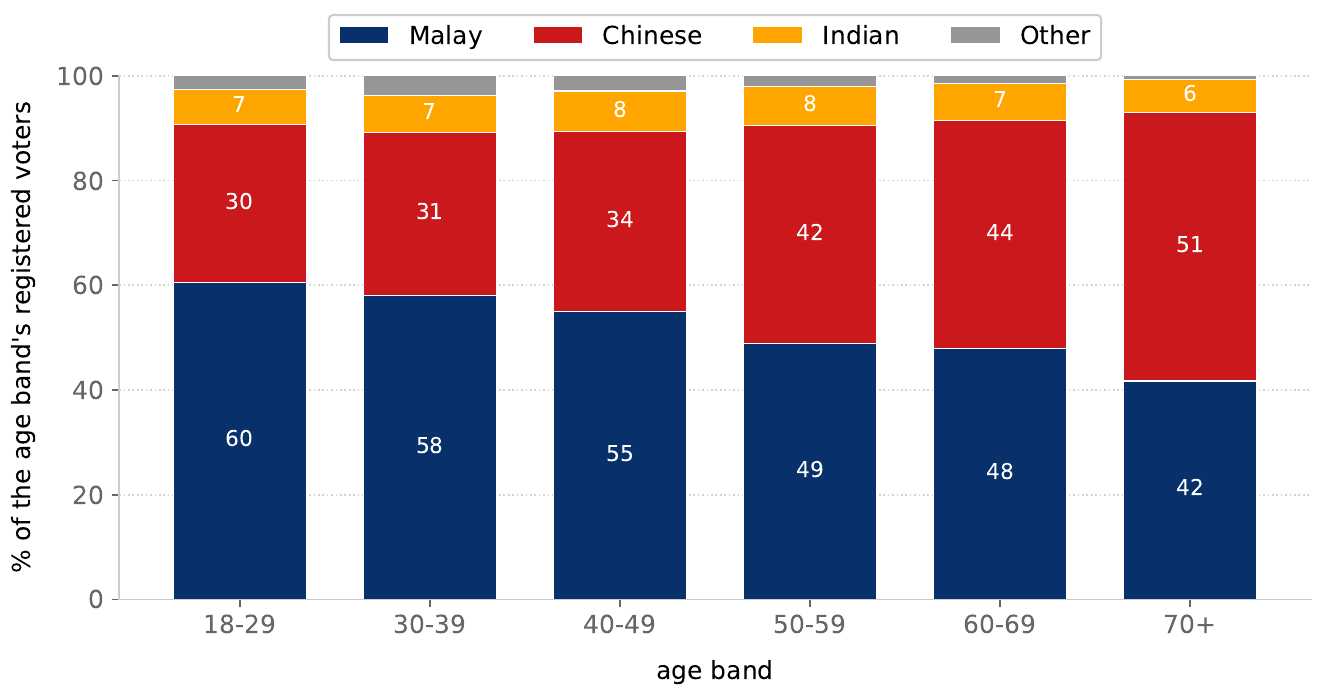}
	\label{fig:agecomp}
\end{figure}

One further feature of the joint distribution shapes interpretation rather than identification: ethnicity and age are strongly associated. Johor's Chinese electorate is much older than its Malay electorate, so each age band has a different ethnic mix (Figure~\ref{fig:agecomp}): the youngest band is 60\% Malay, the oldest 42\% Malay and 51\% Chinese. This association both confounds marginal age patterns (Section~\ref{sec:findings-age}) and causes models that estimate age separately from ethnicity to fail validation (Section~\ref{sec:strategy-groundtruth}).

Each saluran therefore forms a contingency table with exact demographic and electoral margins but an unobserved interior---the number of registered Chinese voters in their sixties who cast a BN ballot, say. Recovering that interior is the classic ecological inference problem \cite{robinson1950,king1997}, here repeated across 14,222 saluran-election tables.

Finally, we exclude early-voting salurans and postal ballots, because we do not have a reliable method to derive the total electorate for these groups, particularly since many individuals originally designated as early voters (primarily police, military, and their spouses) may have subsequently applied to vote via post. The excluded bloc is under 3.5\% of ballots, and including the early salurans moves no statewide estimate by more than about one point.

\section{Empirical strategy}
\label{sec:strategy}

Because EI estimates depend on assumptions about the hidden interior \cite{robinson1950,freedman1998,cho1998}, we build and test the specification in five steps: ecological regression, a bounded link, exact accounting reconciliation, held-out validation, and the estimand-specific error that validation yields.

For group $g$ and outcome $c$---turning out, or voting for a coalition---the estimand is
\begin{equation}
	b_{gc} \;=\; \frac{\text{registered voters in } g \text{ who did } c}{\text{registered voters in } g}.
	\label{eq:estimand}
\end{equation}

The denominator is registration, not turnout, deliberately: staying home is treated as one of the things a voter can do, so a group's turnout and its party rates share one denominator, add up, and keep mobilisation and persuasion inside one accounting frame---which matters when turnout moves by twenty points between elections. Throughout, we call the register the \emph{electorate} and reserve \emph{ballots} for votes actually cast: a group's rates as shares of its electorate sum across outcomes, and its shares of ballots follow by dividing through by turnout.

In every model below, a saluran's rate is the average of its groups' rates, weighted by composition,
\begin{equation}
	p_{ic} \;=\; \sum_g x_{ig}\, b_{igc},
	\label{eq:identity}
\end{equation}
where $p_{ic}$ is saluran $i$'s observed rate of outcome $c$, $x_{ig}$ is group $g$'s share of saluran $i$'s electorate, and $b_{igc}$ is group $g$'s unobserved rate of $c$ in that saluran. Both $x$ and $p$ are observed. Within an EI framework, one needs to impose assumptions in the identity in order to estimate $b$; we construct these carefully below.

One boundary should be drawn before any estimation. Equation~\eqref{eq:estimand} identifies group rates, and changes in those rates, but not individual transition matrices. Equal and opposite coalition movements strongly imply net movement between them, since the arithmetic rules out rival aggregate flows that did not move, but offsetting churn beneath the net figures is invisible to every method used here. We therefore describe net, not individual, movement throughout.

\subsection{Step 1: Classic ecological regression}
\label{sec:strategy-lpm}

The classical estimator assumes each group behaves the same everywhere, $b_{igc}=b_{gc}$, and reads the rates off equation~\eqref{eq:identity} as regression coefficients \cite{goodman1953}. The mechanism at its heart carries every method in this paper. Suppose one stream is 90\% Malay and gives BN 55\%, another is 30\% Malay and gives BN 25\%. If Malay voters have one rate $m$ and everyone else one rate $n$, these are two equations in two unknowns, and they solve exactly: $m=60\%$, $n=10\%$. Group rates fall out of aggregates alone, because streams with different mixes are different blends of the same unknowns. With 4,889 streams the solution becomes a weighted least-squares fit of
\begin{equation}
	p_{ic} \;=\; \sum_g x_{ig}\, b_{gc} + \varepsilon_{ic},
	\tag{M1}\label{eq:m1}
\end{equation}
with saluran electorates $V_i$ as weights, no intercept, and $\varepsilon_{ic}$ the residual; the coefficients \emph{are} the group rates \cite{goodman1953}, \emph{assuming the same} $m$ everywhere.

Six of the 120 group-outcome estimates (Table~\ref{tab:m1}) across the three elections fall outside their deterministic bounds, the worst putting PN support among Indian voters at $-5.2\%$ at SE-16. The failures concentrate where small parties meet groups spread thinly across the state, requiring extrapolation beyond the observed compositional support. These clipped estimates also reproduce a published breakdown of these elections whose estimator is otherwise undocumented.

\begin{table}[!htb]
\caption{Goodman regression (M1) by ethnicity: the constant-rate estimates, as fitted}
\label{tab:m1}
\centering
\footnotesize\setlength{\tabcolsep}{4pt}
\begin{tabular}{ll wc{3.4em} *{3}{wc{2.7em}} @{\hspace{10pt}} *{3}{wc{2.7em}}}
\toprule
 & & \multicolumn{4}{c}{\% of the group's electorate} & \multicolumn{3}{c}{\% of the group's ballots} \\
\cmidrule(lr){3-6}\cmidrule(lr){7-9}
 & & Turnout & BN & PH & PN & BN & PH & PN \\
\midrule
\multirow{3}{*}{\textbf{Malay}} & SE-15 & 63.7 & 36.8 & 2.5 & 21.8 & 57.7 & 4.0 & 34.3 \\
 & GE-15 & 78.6 & 37.7 & 6.2 & 33.7 & 47.9 & 7.9 & 43.0 \\
 & SE-16 & 75.9 & 63.5 & 3.3 & \textcolor{red}{7.9} & 83.8 & 4.3 & \textcolor{red}{10.4} \\
\addlinespace[2pt]
\multirow{3}{*}{\textbf{Chinese}} & SE-15 & 43.9 & 4.3 & 32.5 & 1.0 & 9.9 & 74.0 & 2.2 \\
 & GE-15 & 68.9 & 4.6 & 62.6 & 0.7 & 6.7 & 90.9 & 1.0 \\
 & SE-16 & 61.9 & 7.0 & 52.7 & \textcolor{red}{$-$0.4} & 11.3 & 85.3 & \textcolor{red}{$-$0.6} \\
\addlinespace[2pt]
\multirow{3}{*}{\textbf{Indian}} & SE-15 & 30.9 & 12.2 & 8.1 & 1.5 & 39.4 & 26.3 & 4.7 \\
 & GE-15 & 67.2 & 5.5 & 62.4 & \textcolor{red}{$-$1.1} & 8.2 & 92.9 & \textcolor{red}{$-$1.7} \\
 & SE-16 & 50.5 & 30.2 & 18.0 & \textcolor{red}{$-$5.2} & 59.9 & 35.6 & \textcolor{red}{$-$10.4} \\
\addlinespace[2pt]
\multirow{3}{*}{\textbf{Other}} & SE-15 & 35.9 & 12.8 & 1.3 & 17.0 & 35.7 & 3.5 & 47.3 \\
 & GE-15 & 62.4 & \textcolor{red}{0.9} & 18.7 & 42.7 & \textcolor{red}{1.4} & 30.0 & 68.5 \\
 & SE-16 & 60.0 & 48.7 & 6.8 & \textcolor{red}{$-$3.5} & 81.3 & 11.3 & \textcolor{red}{$-$5.9} \\
\addlinespace[2pt]
\bottomrule
\end{tabular}\par\vspace{3pt}
{\footnotesize \textcolor{red}{red}: outside the deterministic bounds of Table~\ref{tab:bounds}}
\end{table}

\subsection{Step 2: Link function}
\label{sec:strategy-link}

Keeping rates between zero and one requires a link. We keep the mixture of equation~\eqref{eq:identity}, write each group's rate as a transformed coefficient, and fit by binomial maximum likelihood on the outcome counts---$Y_{ic}$, the number of saluran $i$'s electorate of $V_i$ who did $c$:
\begin{align}
	Y_{ic} & \;\sim\; \operatorname{Binomial}\!\Big(V_i,\; \sum_g x_{ig}\operatorname{expit}(\beta_{gc})\Big),
	\tag{M2, logit}\label{eq:m2}                                                                               \\
	Y_{ic} & \;\sim\; \operatorname{Binomial}\!\Big(V_i,\; \sum_g x_{ig}\,\Phi(\beta_{gc})\Big),
	\tag{M3, probit}\label{eq:m3}
\end{align}
so the group rate is $b_{gc}=\operatorname{expit}(\beta_{gc})$ under M2 and $\Phi(\beta_{gc})$ under M3. The transformation applies to the group rates; the saluran remains their observed composition-weighted mixture. Fitting $\operatorname{expit}(\sum_g x_{ig}\beta_g)$ instead would not yield interpretable group rates.

Logit and probit estimates differ by at most 0.030 points, so the link choice is immaterial. We retain the logit because the prior below and the odds-ratio-preserving raking of Step 3 are naturally expressed on that scale.

One corner needs handling: PN is almost absent among non-Malays, so the unpenalised likelihood sends those coefficients to $-\infty$. A weakly informative $N(0,2.5^2)$ prior on each $\beta$ \cite{gelman2008} keeps the corner finite while changing identified estimates by at most 0.034 points. Statewide, one estimate of 120 now sits outside its deterministic bounds, by 0.025 points. That having been said, an estimate can respect its statewide bounds and still be impossible in individual streams, which is what the next step establishes and fixes.

\subsection{Step 3: Accounting constraint}
\label{sec:strategy-rake}

Before any modelling, counting alone rules things out. In a stream of 500 voters, 300 of them Malay, where BN won 400 votes, the 200 non-Malays can supply at most 200 of them: at least 200 and at most 300 BN votes are Malay, with nothing assumed about behaviour. In general, for a saluran with electorate $V$, group size $N_g$ and outcome total $Y_c$, the number of group-$g$ members among those $Y_c$---call it $y_{gc}$---satisfies
\begin{equation}
	\max(0,\; Y_c - (V - N_g)) \;\le\; y_{gc} \;\le\; \min(Y_c,\, N_g).
	\label{eq:bounds}
\end{equation}
These sharp bounds \cite{duncandavis1953} rest only on the observed margins: the true cell count must lie within them. Their width therefore measures how much each estimate is constrained by arithmetic rather than modelling \cite{chogaines2004}, and they bind hardest where a stream is dominated by one group---which is why Johor's near-pure streams matter so much in Step 4. Table~\ref{tab:bounds} reports them statewide, for every group and election: any account of these elections must live inside these intervals, regardless of the model used to estimate it.

\begin{table}[!htb]
\caption{The deterministic accounting bounds, statewide: every estimate must lie inside them}
\label{tab:bounds}
\centering
\footnotesize\setlength{\tabcolsep}{5pt}
\begin{tabular}{ll cccc}
\toprule
 & & Turnout & BN & PH & PN \\
\midrule
\multirow{3}{*}{\textbf{Malay}} & SE-15 & [37.7, 79.2] & [16.0, 41.2] & [0.3, 20.1] & [7.4, 23.5] \\
 & GE-15 & [58.2, 91.8] & [16.4, 40.8] & [1.6, 37.6] & [14.0, 36.1] \\
 & SE-16 & [52.7, 89.9] & [34.3, 72.0] & [0.5, 28.7] & [2.2, 6.8] \\
\addlinespace[2pt]
\multirow{3}{*}{\textbf{Chinese}} & SE-15 & [18.1, 76.5] & [0.5, 31.6] & [5.8, 36.5] & [0.0, 19.3] \\
 & GE-15 & [46.9, 94.3] & [0.6, 30.9] & [20.8, 76.9] & [0.0, 26.4] \\
 & SE-16 & [36.6, 91.2] & [1.0, 49.8] & [12.6, 59.4] & [0.0, 5.6] \\
\addlinespace[2pt]
\multirow{3}{*}{\textbf{Indian}} & SE-15 & [1.0, 99.4] & [0.2, 82.8] & [0.0, 72.1] & [0.0, 58.5] \\
 & GE-15 & [3.9, 99.9] & [0.1, 79.4] & [0.3, 97.7] & [0.0, 68.8] \\
 & SE-16 & [1.7, 99.8] & [0.3, 96.5] & [0.0, 91.1] & [0.0, 14.8] \\
\addlinespace[2pt]
\multirow{3}{*}{\textbf{Other}} & SE-15 & [2.3, 99.1] & [1.1, 95.6] & [0.1, 68.4] & [0.4, 88.9] \\
 & GE-15 & [3.5, 99.2] & [1.3, 93.4] & [0.4, 89.4] & [0.2, 94.6] \\
 & SE-16 & [2.5, 99.3] & [1.2, 98.5] & [0.1, 83.0] & [0.0, 34.5] \\
\addlinespace[2pt]
\midrule
\multirow{3}{*}{\textbf{18--29}} & SE-15 & [38.9, 63.3] & [13.2, 29.9] & [7.5, 16.5] & [8.2, 20.8] \\
 & GE-15 & [63.5, 83.0] & [10.9, 26.2] & [21.9, 38.5] & [17.1, 33.7] \\
 & SE-16 & [54.4, 77.1] & [31.5, 54.4] & [11.8, 25.4] & [1.6, 6.0] \\
\addlinespace[2pt]
\multirow{3}{*}{\textbf{30--39}} & SE-15 & [23.0, 71.6] & [5.9, 38.3] & [3.0, 21.1] & [3.6, 27.9] \\
 & GE-15 & [48.9, 89.6] & [6.0, 37.8] & [11.9, 46.7] & [7.4, 41.5] \\
 & SE-16 & [42.9, 85.4] & [20.9, 63.2] & [7.2, 33.2] & [0.7, 8.9] \\
\addlinespace[2pt]
\multirow{3}{*}{\textbf{40--49}} & SE-15 & [25.0, 78.9] & [4.6, 40.7] & [3.8, 27.7] & [2.5, 28.4] \\
 & GE-15 & [51.5, 93.4] & [5.0, 42.0] & [14.6, 54.8] & [4.3, 38.5] \\
 & SE-16 & [44.5, 89.7] & [17.7, 64.0] & [7.2, 39.0] & [0.7, 9.8] \\
\addlinespace[2pt]
\multirow{3}{*}{\textbf{50--59}} & SE-15 & [30.6, 87.0] & [4.7, 47.6] & [5.2, 34.2] & [1.6, 27.7] \\
 & GE-15 & [55.5, 96.2] & [5.4, 48.2] & [16.4, 59.6] & [2.5, 33.7] \\
 & SE-16 & [48.1, 94.0] & [14.4, 64.5] & [10.3, 49.1] & [0.4, 10.3] \\
\addlinespace[2pt]
\multirow{3}{*}{\textbf{60--69}} & SE-15 & [26.6, 90.1] & [5.0, 53.4] & [5.9, 36.8] & [0.9, 23.1] \\
 & GE-15 & [48.4, 96.2] & [6.7, 55.0] & [13.9, 59.9] & [1.2, 26.2] \\
 & SE-16 & [46.6, 95.8] & [11.6, 66.2] & [11.0, 53.2] & [0.2, 8.2] \\
\addlinespace[2pt]
\multirow{3}{*}{\textbf{70+}} & SE-15 & [19.9, 79.9] & [3.4, 49.3] & [3.5, 30.3] & [0.0, 19.6] \\
 & GE-15 & [36.9, 89.3] & [4.5, 52.0] & [9.2, 51.7] & [0.0, 22.6] \\
 & SE-16 & [40.4, 87.2] & [9.4, 56.8] & [9.1, 44.5] & [0.0, 6.6] \\
\addlinespace[2pt]
\bottomrule
\end{tabular}
\end{table}

M2 still violates the constraint locally: 18.5\% of saluran-level cells across the three elections are impossible under M2's own estimates, rising to 23.9\% at SE-16, and its implied statewide BN total at SE-16 misses the observed count by 2,501 ballots. A decomposition that fails either margin is not coherent.

The constraint has to be imposed where it lives: the saluran. We therefore rake each saluran's group-by-choice table to its observed margins, starting from M2's statewide rates. Iterative proportional fitting yields the unique non-negative table matching both margins while preserving the starting table's odds ratios \cite{demingstephan1940,irelandkullback1968}---structure-preserving estimation \cite{purcellkish1980}, the standard small-area tool for this shape of problem: margins known, interior borrowed. Because the bounds follow from the margins, imposing the margins imposes the bounds in every stream. We verify that there are no violations anywhere, that margins are matched to $2\times10^{-13}$, and that every statewide total is reproduced exactly.

Raking also splits the work sensibly between data and model, since near-pure salurans are determined largely by their margins while mixed ones inherit more from the model. What it buys, though, is coherence rather than accuracy: scored against the held-out outcomes of Step 4, the raked estimates miss by 4.15 points on average against 4.12 for raw M2. Accuracy improves only in Step 4, when the rows are redefined.

\subsection{Step 4: validation against observed behaviour}
\label{sec:strategy-groundtruth}

Because the Commission sorts voters into streams by age, a large share of streams contain, near enough, a single age band---and in such a stream the published result is very close to that band's own turnout and vote. That supports a held-out test: fit the model on \emph{mixed} salurans only, predict the near-pure ones, and score the predictions against what those streams actually did. The idea has two lineages: homogeneous-precinct analysis \cite{grofman1992}, and the small body of work validating EI wherever an administrative accident recorded the truth \cite{hudson2010,plesciadesio2018,park2014}.

A 95\%-pure stream does not literally reveal the dominant group's rate, because the remaining 5\% still contribute to the result. We therefore use the exact interval permitted by the margins as the target, applying equation~\eqref{eq:bounds} to the pure subset. At 95\% purity these intervals average 0.7 points wide for the age bands and 2.3 for the validated ethnic groups---narrower for age because the sorting rule leaves many streams exactly one band, while ethnically pure streams only just clear the threshold---and a prediction scores zero whenever it falls inside the interval, so no credit is taken for its width.

At the 95\% threshold Johor supplies 1,066--1,122 age-pure salurans per election, and 735--754 ethnically pure ones. Two caveats bound what the test can show. The first is selection: because pure and mixed streams may occupy different localities, we score both an unrestricted holdout (design A) and one restricted to districts containing both types (design B). The restricted design produces slightly smaller joint-model misses---0.76 against 0.97 points at SE-16---while leaving the decisive 70+ comparison unchanged, the separate model missing that band by 10.0 points under A and 10.7 under B. We therefore report the unrestricted, more conservative design; the full grid is in the replication output.

The second caveat is coverage, which is uneven. At SE-16 the age-pure share falls from 56.0\% of the 18--29 electorate to 4.4\% of the over-70s, while ethnic holdouts cover 24.5\% of Malays, 2.8\% of Chinese, and no usable Indian or Other stratum at any threshold. The validation is therefore strongest for age and Malays, weaker for Chinese, and silent exactly where the estimates are already flagged as non-citable. Selection also runs in opposite directions across the two dimensions---age-pure streams sit in larger districts, ethnically pure ones in districts half the size with turnout 9.3 points higher---so no single behavioural confound easily explains the passing results in both.

The model which estimates ethnicity and age separately has a mean miss of 4.93 points across the held-out age outcomes, reaching 19 points on BN support among the over-70s. The reason is intuitive to understand, since it is a function of the sorting rule. Johor's Chinese are older on average, so the oldest stream in a district is disproportionately the Chinese stream: the electorate of age-pure 70+ salurans is 86\% Chinese, against 51\% for the 70+ electorate as a whole. A model with one rate per age band cannot tell a 70-year-old Chinese voter from a 70-year-old Malay voter, and is wrong in precisely the way the administrative design guarantees.

Because the roll records ethnicity and birth year for the same person, every saluran's 24 ethnicity-by-age margins are counted exactly rather than modelled. Replacing the 10 separate ethnic and age rows with these 24 joint cells---without changing the likelihood, the prior or the raking---cuts the mean age miss from 4.93 to 0.98 points, the worst 70+ miss from 19 points to 5, and the ethnic cells from 1.79 to 1.39. However, it does not fix the under-identification of Indian and Other voters.

One comparison puts this in context. The hierarchical Bayesian R$\times$C model of Rosen et al.\cite{rosen2001}, as implemented in eiPack \cite{eipack2007}, run on the same held-out streams, gives a mean miss of 5.49 points against our 4.88 on the same specification. The two are indistinguishable and fail in the same places. In this problem, gains come from richer rows, not richer likelihoods.

\subsection{Step 5: what the error bars measure}
\label{sec:strategy-errors}

Conventional sampling errors do not address the dominant uncertainty here. The margins are a finite population that was counted---2,703,175 registered voters and every ballot they cast---so what constrains us is identification: which member of the admissible set we report depends on a behavioural assumption the data cannot test, and being wrong about it produces \emph{bias}, not noise, which 2.7 million voters do nothing to shrink \cite{manski1995}. The distinction is substantive, not pedantic: Indian turnout at SE-16 has a bootstrap standard error under three points on a quantity whose deterministic bounds run from 1.7\% to 99.8\%. We therefore report held-out prediction error rather than sampling precision.

Step 4 has already produced that measurement. For each group, election and estimand, the $\pm$ we print is the statewide distance between prediction and what the corresponding held-out streams actually did. Errors are estimand-specific because turnout and the three coalition outcomes are not predicted equally well---Chinese PN support is near zero with nowhere to go, while Malay BN support at GE-15 is the hardest quantity in the paper---so an average across the four would overstate three and understate the fourth.

Ballot shares are the exception, and require propagation rather than direct scoring. Scoring a ratio against its own bounds flatters it twice, because the admissible interval of a ratio is roughly twice as wide as its components' and the component errors can cancel: at SE-16 the model undershoots both Malay turnout (74.3 against permitted bounds $[76.8, 78.8]$) and Malay BN votes (62.8 against $[63.7, 65.6]$), yet their ratio sits inside the wider interval and records a miss of zero. The ballot-share bars therefore inherit their components' errors without assuming cancellation, $E_v = (E_{\text{party}} + s\,E_{\text{turnout}})/t$, where $t$ is the group's turnout and $s$ the ballot share itself---the first-order propagation of both component errors through the ratio $s = b_{\text{party}}/t$---and we keep the direct measurement only where it is larger.

The error is formed at the statewide scale of the reported estimand. Saluran-level misses, which average 5.8 points, measure local heterogeneity rather than error in the statewide aggregate.

One statewide miss is a single number, and which corners of Johor happen to contain validatable streams is arbitrary, so we resample. Let $H_g$ be the near-pure streams whose dominant group is $g$, and $D_g$ the polling districts containing them. A replicate draws $|D_g|$ districts with replacement and keeps every held-out stream in each, so streams in one place move together. Within a replicate, the target interval $[L, U]$ and the prediction $\hat p$ are aggregated over the drawn streams, weighting by $N_{ig}$, the group's registered electorate in stream $i$; $Y_i$ and $V_i$ are stream $i$'s outcome total and electorate, and $\hat p_i$ the model's predicted rate for the group there:
\begin{equation}
	L = \frac{\sum_i \max\{0,\, Y_i - (V_i - N_{ig})\}}{\sum_i N_{ig}}, \qquad
	U = \frac{\sum_i \min\{Y_i,\, N_{ig}\}}{\sum_i N_{ig}}, \qquad
	\hat p = \frac{\sum_i N_{ig}\, \hat p_i}{\sum_i N_{ig}},
	\label{eq:aggmiss}
\end{equation}
giving a miss of $\max\{L - \hat p,\, \hat p - U,\, 0\}$---zero whenever the prediction is admissible. The reported error is the root mean square of that miss across 2,000 replicates: a root mean square over replicates of one statewide quantity, not over estimands or streams. Table~\ref{tab:errors} gives the full set.

\begin{table}[!htb]
\caption{Held-out validation error by group, election and estimand, in percentage points}
\label{tab:errors}
\centering\footnotesize\setlength{\tabcolsep}{4pt}
\begin{tabular}{ll cccc @{\hspace{6pt}} ccc}
\toprule
 & & \multicolumn{4}{c}{\% of the group's \textbf{electorate}} & \multicolumn{3}{c}{\% of the group's \textbf{ballots}} \\
\cmidrule(lr){3-6}\cmidrule(lr){7-9}
 & & Turnout & BN & PH & PN & BN & PH & PN \\
\midrule
\multirow{3}{*}{\textbf{Malay}} & SE-15 & 3.66 & 5.44 & 0.74 & 0.92 & 11.92 & 1.41 & 3.47 \\
 & GE-15 & 1.59 & 6.06 & 1.96 & 2.46 & 8.69 & 2.68 & 4.03 \\
 & SE-16 & 2.47 & 1.01 & 0.41 & 0.69 & 4.11 & 0.70 & 1.21 \\
\addlinespace[2pt]
\multirow{3}{*}{\textbf{Chinese}} & SE-15 & 3.55 & 0.13 & 1.28 & $<$0.01 & 1.16 & 9.05 & 0.17 \\
 & GE-15 & 1.60 & 0.22 & 0.30 & 0.01 & 0.48 & 2.55 & 0.04 \\
 & SE-16 & 2.19 & 0.26 & 0.30 & $<$0.01 & 0.84 & 3.57 & 0.01 \\
\addlinespace[2pt]
\midrule
\multirow{3}{*}{\textbf{18--29}} & SE-15 & 0.20 & 1.72 & 0.68 & 0.25 & 3.50 & 1.40 & 0.59 \\
 & GE-15 & 1.51 & 3.10 & 2.94 & 1.49 & 4.72 & 4.79 & 2.72 \\
 & SE-16 & 0.83 & 0.85 & 0.24 & 0.98 & 2.10 & 0.70 & 1.55 \\
\addlinespace[2pt]
\multirow{3}{*}{\textbf{30--39}} & SE-15 & 0.53 & 0.25 & 0.37 & 0.13 & 1.03 & 1.05 & 0.62 \\
 & GE-15 & 0.34 & 0.47 & 0.18 & 0.35 & 0.82 & 0.44 & 0.66 \\
 & SE-16 & 0.19 & 0.46 & 0.34 & 0.63 & 0.89 & 0.59 & 0.98 \\
\addlinespace[2pt]
\multirow{3}{*}{\textbf{40--49}} & SE-15 & 1.23 & 1.12 & 1.45 & 0.37 & 3.06 & 3.40 & 1.32 \\
 & GE-15 & 1.42 & 0.77 & 1.80 & 0.34 & 1.55 & 3.24 & 0.95 \\
 & SE-16 & 0.96 & 0.42 & 1.12 & 1.17 & 1.45 & 2.07 & 1.81 \\
\addlinespace[2pt]
\multirow{3}{*}{\textbf{50--59}} & SE-15 & 1.73 & 1.19 & 2.11 & 0.29 & 2.99 & 4.23 & 1.05 \\
 & GE-15 & 0.24 & 2.05 & 1.21 & 0.28 & 2.65 & 1.65 & 0.42 \\
 & SE-16 & 0.39 & 1.47 & 0.79 & 0.46 & 2.23 & 1.26 & 0.65 \\
\addlinespace[2pt]
\multirow{3}{*}{\textbf{60--69}} & SE-15 & 2.22 & 2.22 & 1.75 & 0.74 & 4.93 & 4.04 & 1.64 \\
 & GE-15 & 0.99 & 1.50 & 0.75 & 0.60 & 2.34 & 1.49 & 0.91 \\
 & SE-16 & 1.26 & 2.23 & 0.50 & 0.95 & 3.64 & 1.33 & 1.27 \\
\addlinespace[2pt]
\multirow{3}{*}{\textbf{70+}} & SE-15 & 3.68 & 3.14 & 1.13 & 0.24 & 10.88 & 4.78 & 1.32 \\
 & GE-15 & 2.88 & 2.29 & 1.70 & 0.17 & 6.29 & 5.07 & 0.68 \\
 & SE-16 & 4.09 & 2.32 & 5.59 & 0.32 & 7.39 & 12.19 & 0.75 \\
\addlinespace[2pt]
\bottomrule
\end{tabular}
\end{table}

The resulting error is empirical, election-specific and estimand-specific: Malay BN support carries $\pm6.1$ points at GE-15 and $\pm1.0$ at SE-16, because the green wave put Malay voters somewhere the constancy assumption fits them badly, so pooling across elections would contradict the paper's own subject. It is also conservative twice over, since the scored model is fitted on mixed salurans only and the unrestricted design leaves locality differences inside the measured error. Against that, it transports performance from near-pure into mixed streams---the principal substantive assumption the bars rest on, since the units in which behaviour is checkable are also those most tightly constrained by arithmetic. The bars describe typical error, not confidence intervals, and carry no nominal coverage.

Results are sensitive to the purity threshold in the expected direction: the mean aggregate miss runs 0.51 points at 90\%, 1.14 at 95\% and 1.74 at 98\%, because a stricter threshold narrows the permitted interval and so catches the same prediction error more often. We use 95\% as a balance between the sharpness of the target and the size of the strata.

\subsubsection*{Movements, and why they survive}

Most claims in Section~\ref{sec:findings} concern movement between elections, so two validation errors must be combined: the error on a movement is the sum of the measured errors of the same quantity at the two elections being compared,
\begin{equation}
	E_{\Delta} = E_{\text{from}} + E_{\text{to}}.
	\label{eq:combine}
\end{equation}
Adding assumes the worst case, that both misses point the same way. The root sum of squares is smaller but assumes independence, which we have no grounds to assert: the same geography, sorting rule and estimator sit behind both elections, and a misspecification that persists is exactly the kind that will not cancel. Every headline movement in Section~\ref{sec:findings} exceeds this combined error, and the two null claims---stable Malay turnout between GE-15 and SE-16, and stable Malay BN support across the two 2022 elections---do not, so both enter only as statements that something did \emph{not} move.

\subsubsection*{Groups we do not report}

Indian and Other estimates carry no error bar, because the bar is the variability of the held-out miss across polling districts and Johor supplies no usable near-pure stratum for either group. The classification is not threshold-sensitive: at 95\% purity every reported group spans at least 19 districts---Malays 217--224, the youngest band 310--355---whereas Indian voters have no near-pure saluran at any election and Other voters at most five, spanning one to four districts and holding at most 1.8\% of the group. Any threshold between 5 and 19 therefore classifies every group identically; we use ten. The over-70 stratum spans only 19 to 29 districts, short of the thirty clusters conventionally held to be needed for cluster-based variance estimation \cite{cameronmiller2015,cameronetal2008}, so its bars are the least reliable in Table~\ref{tab:errors}; we report them rather than drop the band carrying the paper's clearest life-cycle finding, on the strength of a convention designed for hypothesis tests we do not run. Which groups a design can validate is an artefact of the particular electorate, not a general rule.

What \emph{can} be said about the two groups is the bounds of Table~\ref{tab:bounds}, which need no model. For Indian turnout at SE-16 the margins permit anything from 1.7\% to 99.8\%, and for Indian BN support 0.3\% to 96.5\%: that is the absence of information, not a precision problem a better estimator could narrow. The bounds are per cell, not per group---for instance, Indian PN support is confined to $[0.0, 14.8]$.

\subsection{The preferred specification}
\label{sec:strategy-preferred}

Table~\ref{tab:models} summarises the progression. Link choice and raking barely affect held-out accuracy---the linear model, the logit and raked M2 land within 0.03 points of one another---while raking alone restores feasibility in every stream and exact margins. Accuracy improves only in the last row, when the ten separate ethnic and age rows are replaced by 24 joint cells. The estimator was never the binding constraint; the row definition was.

\begin{table}[!htb]
	\caption{The specifications compared}
	\label{tab:models}
	\centering
	\footnotesize
	\setlength{\tabcolsep}{3pt}
	\begin{tabular}{lccccc}
		\toprule
		                                                          & Rates in     & Possible     & Totals       & \multicolumn{2}{c}{Mean $|$miss$|$}                 \\
		\cmidrule(lr){5-6}
		Specification                                             & $[0,1]$      & everywhere   & exact        & Age                                 & Ethnicity     \\
		\midrule
		Bivariate scatterplot                                     & no           & no           & no           & 5.37                                & 2.57          \\
		M1 Goodman LPM                                            & no           & no           & no           & 4.95                                & 1.73          \\
		M2 ecological logit                                       & yes          & no           & no           & 4.93                                & 1.70          \\
		M3 probit                                                 & yes          & no           & no           & 4.93                                & 1.70          \\
		Raked M2, 4 ethnicity $+$ 6 age $=$ 10 rows               & yes          & yes          & yes          & 4.93                                & 1.79          \\
		\textbf{Raked M2, 4 ethnicity $\times$ 6 age $=$ 24 rows} & \textbf{yes} & \textbf{yes} & \textbf{yes} & \textbf{0.98}                       & \textbf{1.39} \\
		\bottomrule
	\end{tabular}
\end{table}

We therefore report raked M2 on the 24 joint rows throughout: every cell is feasible, every margin is exact, and the specification passes the held-out test. Among validated groups its marginals differ from the separate model's by at most 1.7 points---the numbers barely move, but their provenance changes from a specification that fails a real-data test to one that passes it.

The bivariate row is the method implicit in every scatterplot-with-a-trendline published after a Malaysian election: fitting vote share against a group's electorate share and reading the line off at 100\% is algebraically identical to two-group ecological regression. It performs tolerably where near-pure salurans exist, landing within 1--3 points of the full model, but where they do not it extrapolates far beyond the data---Indian BN support of $-23.1\%$ at GE-15, with ten of the 24 Indian and Other extrapolations arithmetically impossible.

\subsection{What the estimates cannot do}
\label{sec:strategy-limits}

Six limits bound what follows. The estimand is a group aggregate, not an individual transition. The validation checks the behavioural assumption only where near-pure salurans exist---directly for the age bands and Malays, partially for Chinese, not at all for Indian and Other. The error bars are measured in near-pure salurans and carried to the rest of the electorate, which assumes the streams we cannot validate are no harder than the ones we can---the assumption most likely to flatter us. The bars are typical errors, not confidence intervals. All results describe the ordinary polling-day electorate. And nothing here identifies \emph{why} groups moved.

One further limit is about Johor rather than method: its electoral geography is unusually favourable to EI, with extreme Malay--Chinese compositional variation across 4,900 small units. A more evenly mixed state would give wider bounds and weaker identification---and the same pipeline, applied there, would report that in the validation error, or decline to report at all.

\section{Findings and discussion}
\label{sec:findings}

Table~\ref{tab:results} reports the preferred estimates for all three elections on both scales, as shares of each group's electorate and of its ballots; Table~\ref{tab:joint} gives the SE-16 joint crosstab and Table~\ref{tab:errors} the error bars. Displayed $\pm$ values are held-out validation errors, not confidence intervals, and point estimates without them---Indian and Other---are not used substantively below.

This section reports what moved, by how much and among whom; the explanations offered for \emph{why} remain interpretive.

\begin{table}[!htb]
\caption{Turnout and coalition support in Johor, 2022--2026, by ethnicity and age}
\label{tab:results}
\centering{\footnotesize (point estimates without error bars are not citable)}\par\vspace{3pt}
\footnotesize\setlength{\tabcolsep}{3.0pt}
\begin{tabular}{ll cccc @{\hspace{6pt}} ccc}
\toprule
 & & \multicolumn{4}{c}{\% of the group's \textbf{electorate}} & \multicolumn{3}{c}{\% of the group's \textbf{ballots}} \\
\cmidrule(lr){3-6}\cmidrule(lr){7-9}
 & & Turnout & BN & PH & PN & BN & PH & PN \\
\midrule

\multirow{3}{*}{\textbf{Malay}} & SE-15 & \cellcolor[HTML]{4090C5}\textcolor{white}{63.1\,{\tiny$\pm$}3.7} & \cellcolor[HTML]{A3CCE3}36.0\,{\tiny$\pm$}5.4 & \cellcolor[HTML]{F2F8FD}2.6\,{\tiny$\pm$}0.7 & \cellcolor[HTML]{CCDFF1}22.0\,{\tiny$\pm$}0.9 & \cellcolor[HTML]{539ECD}\textcolor{white}{57.1\,{\tiny$\pm$}11.9} & \cellcolor[HTML]{EFF6FC}4.1\,{\tiny$\pm$}1.4 & \cellcolor[HTML]{A6CEE4}34.8\,{\tiny$\pm$}3.5 \\
 & GE-15 & \cellcolor[HTML]{1A68AE}\textcolor{white}{78.3\,{\tiny$\pm$}1.6} & \cellcolor[HTML]{A1CBE2}36.5\,{\tiny$\pm$}6.1 & \cellcolor[HTML]{EAF3FB}6.3\,{\tiny$\pm$}2.0 & \cellcolor[HTML]{A8CEE4}34.5\,{\tiny$\pm$}2.5 & \cellcolor[HTML]{79B5D9}46.7\,{\tiny$\pm$}8.7 & \cellcolor[HTML]{E7F1FA}8.1\,{\tiny$\pm$}2.7 & \cellcolor[HTML]{84BCDB}44.0\,{\tiny$\pm$}4.0 \\
 & SE-16 & \cellcolor[HTML]{206FB4}\textcolor{white}{75.5\,{\tiny$\pm$}2.5} & \cellcolor[HTML]{3D8DC4}\textcolor{white}{64.1\,{\tiny$\pm$}1.0} & \cellcolor[HTML]{F0F6FD}3.6\,{\tiny$\pm$}0.4 & \cellcolor[HTML]{EAF2FB}6.8\,{\tiny$\pm$}0.7 & \cellcolor[HTML]{0D57A1}\textcolor{white}{84.9\,{\tiny$\pm$}4.1} & \cellcolor[HTML]{EEF5FC}4.8\,{\tiny$\pm$}0.7 & \cellcolor[HTML]{E6F0F9}9.0\,{\tiny$\pm$}1.2 \\
\addlinespace[2pt]
\multirow{3}{*}{\textbf{Chinese}} & SE-15 & \cellcolor[HTML]{89BEDC}42.8\,{\tiny$\pm$}3.6 & \cellcolor[HTML]{EEF5FC}4.5\,{\tiny$\pm$}0.1 & \cellcolor[HTML]{B2D2E8}31.3\,{\tiny$\pm$}1.3 & \cellcolor[HTML]{F5FAFE}0.9\,{\tiny$\pm<$}0.1 & \cellcolor[HTML]{E3EEF8}10.4\,{\tiny$\pm$}1.2 & \cellcolor[HTML]{2575B7}\textcolor{white}{73.1\,{\tiny$\pm$}9.1} & \cellcolor[HTML]{F3F8FE}2.1\,{\tiny$\pm$}0.2 \\
 & GE-15 & \cellcolor[HTML]{3181BD}\textcolor{white}{69.0\,{\tiny$\pm$}1.6} & \cellcolor[HTML]{EEF5FC}4.6\,{\tiny$\pm$}0.2 & \cellcolor[HTML]{4090C5}\textcolor{white}{63.1\,{\tiny$\pm$}0.3} & \cellcolor[HTML]{F6FAFF}0.5\,{\tiny$\pm<$}0.1 & \cellcolor[HTML]{EAF2FB}6.6\,{\tiny$\pm$}0.5 & \cellcolor[HTML]{08468B}\textcolor{white}{91.5\,{\tiny$\pm$}2.6} & \cellcolor[HTML]{F6FAFF}0.8\,{\tiny$\pm<$}0.1 \\
 & SE-16 & \cellcolor[HTML]{4695C8}\textcolor{white}{61.0\,{\tiny$\pm$}2.2} & \cellcolor[HTML]{EAF2FB}6.8\,{\tiny$\pm$}0.3 & \cellcolor[HTML]{64A9D3}\textcolor{white}{52.2\,{\tiny$\pm$}0.3} & \cellcolor[HTML]{F7FBFF}0.0\,{\tiny$\pm<$}0.1 & \cellcolor[HTML]{E1EDF8}11.2\,{\tiny$\pm$}0.8 & \cellcolor[HTML]{0B559F}\textcolor{white}{85.6\,{\tiny$\pm$}3.6} & \cellcolor[HTML]{F7FBFF}0.1\,{\tiny$\pm<$}0.1 \\
\addlinespace[2pt]
\multirow{3}{*}{\textbf{Indian}} & SE-15 & 35.3 & 11.9 & 13.0 & 2.4 & 33.7 & 36.8 & 6.7 \\
 & GE-15 & 68.6 & 6.2 & 61.0 & 0.6 & 9.1 & 88.8 & 0.9 \\
 & SE-16 & 55.3 & 27.2 & 17.3 & 0.0 & 49.2 & 31.3 & 0.0 \\
\addlinespace[2pt]
\multirow{3}{*}{\textbf{Other}} & SE-15 & 50.6 & 27.2 & 4.2 & 13.0 & 53.7 & 8.3 & 25.7 \\
 & GE-15 & 62.8 & 22.1 & 13.5 & 25.5 & 35.3 & 21.6 & 40.7 \\
 & SE-16 & 65.7 & 47.7 & 7.9 & 0.1 & 72.6 & 12.0 & 0.1 \\
\addlinespace[2pt]
\midrule
\multirow{3}{*}{\textbf{18--29}} & SE-15 & \cellcolor[HTML]{65AAD4}\textcolor{white}{51.6\,{\tiny$\pm$}0.2} & \cellcolor[HTML]{CDE0F1}21.4\,{\tiny$\pm$}1.7 & \cellcolor[HTML]{E0ECF8}11.4\,{\tiny$\pm$}0.7 & \cellcolor[HTML]{DBE9F6}14.3\,{\tiny$\pm$}0.2 & \cellcolor[HTML]{8DC1DD}41.5\,{\tiny$\pm$}3.5 & \cellcolor[HTML]{CCDFF1}22.2\,{\tiny$\pm$}1.4 & \cellcolor[HTML]{BED8EC}27.7\,{\tiny$\pm$}0.6 \\
 & GE-15 & \cellcolor[HTML]{2373B6}\textcolor{white}{74.2\,{\tiny$\pm$}1.5} & \cellcolor[HTML]{D4E4F4}17.8\,{\tiny$\pm$}3.1 & \cellcolor[HTML]{B5D4E9}30.2\,{\tiny$\pm$}2.9 & \cellcolor[HTML]{C4DAEE}25.7\,{\tiny$\pm$}1.5 & \cellcolor[HTML]{C8DCF0}24.0\,{\tiny$\pm$}4.7 & \cellcolor[HTML]{91C3DE}40.7\,{\tiny$\pm$}4.8 & \cellcolor[HTML]{A8CEE4}34.7\,{\tiny$\pm$}2.7 \\
 & SE-16 & \cellcolor[HTML]{3888C1}\textcolor{white}{66.0\,{\tiny$\pm$}0.8} & \cellcolor[HTML]{89BEDC}42.9\,{\tiny$\pm$}0.9 & \cellcolor[HTML]{D4E4F4}17.7\,{\tiny$\pm$}0.2 & \cellcolor[HTML]{F1F7FD}3.4\,{\tiny$\pm$}1.0 & \cellcolor[HTML]{3B8BC2}\textcolor{white}{64.9\,{\tiny$\pm$}2.1} & \cellcolor[HTML]{C1D9ED}26.9\,{\tiny$\pm$}0.7 & \cellcolor[HTML]{EDF4FC}5.2\,{\tiny$\pm$}1.6 \\
\addlinespace[2pt]
\multirow{3}{*}{\textbf{30--39}} & SE-15 & \cellcolor[HTML]{7AB6D9}46.1\,{\tiny$\pm$}0.5 & \cellcolor[HTML]{D1E2F3}19.5\,{\tiny$\pm$}0.2 & \cellcolor[HTML]{E3EEF9}10.0\,{\tiny$\pm$}0.4 & \cellcolor[HTML]{DCEAF6}13.4\,{\tiny$\pm$}0.1 & \cellcolor[HTML]{8ABFDD}42.3\,{\tiny$\pm$}1.0 & \cellcolor[HTML]{CDDFF1}21.7\,{\tiny$\pm$}1.0 & \cellcolor[HTML]{B9D6EA}29.1\,{\tiny$\pm$}0.6 \\
 & GE-15 & \cellcolor[HTML]{2F7FBC}\textcolor{white}{69.7\,{\tiny$\pm$}0.3} & \cellcolor[HTML]{D0E1F2}20.0\,{\tiny$\pm$}0.5 & \cellcolor[HTML]{C1D9ED}26.7\,{\tiny$\pm$}0.2 & \cellcolor[HTML]{CBDEF1}22.4\,{\tiny$\pm$}0.4 & \cellcolor[HTML]{BAD6EB}28.7\,{\tiny$\pm$}0.8 & \cellcolor[HTML]{9AC8E0}38.3\,{\tiny$\pm$}0.4 & \cellcolor[HTML]{AFD1E7}32.1\,{\tiny$\pm$}0.7 \\
 & SE-16 & \cellcolor[HTML]{3989C1}\textcolor{white}{65.6\,{\tiny$\pm$}0.2} & \cellcolor[HTML]{8CC0DD}41.8\,{\tiny$\pm$}0.5 & \cellcolor[HTML]{D3E4F3}18.3\,{\tiny$\pm$}0.3 & \cellcolor[HTML]{F0F6FD}3.8\,{\tiny$\pm$}0.6 & \cellcolor[HTML]{3E8EC4}\textcolor{white}{63.7\,{\tiny$\pm$}0.9} & \cellcolor[HTML]{BDD7EC}27.9\,{\tiny$\pm$}0.6 & \cellcolor[HTML]{ECF4FB}5.7\,{\tiny$\pm$}1.0 \\
\addlinespace[2pt]
\multirow{3}{*}{\textbf{40--49}} & SE-15 & \cellcolor[HTML]{64A9D3}\textcolor{white}{52.3\,{\tiny$\pm$}1.2} & \cellcolor[HTML]{D0E1F2}20.3\,{\tiny$\pm$}1.1 & \cellcolor[HTML]{DCE9F6}14.0\,{\tiny$\pm$}1.4 & \cellcolor[HTML]{DCEAF6}13.6\,{\tiny$\pm$}0.4 & \cellcolor[HTML]{99C7E0}38.8\,{\tiny$\pm$}3.1 & \cellcolor[HTML]{C1D9ED}26.8\,{\tiny$\pm$}3.4 & \cellcolor[HTML]{C3DAEE}26.0\,{\tiny$\pm$}1.3 \\
 & GE-15 & \cellcolor[HTML]{2070B4}\textcolor{white}{75.1\,{\tiny$\pm$}1.4} & \cellcolor[HTML]{CDE0F1}21.2\,{\tiny$\pm$}0.8 & \cellcolor[HTML]{AACFE5}33.6\,{\tiny$\pm$}1.8 & \cellcolor[HTML]{D1E2F3}19.5\,{\tiny$\pm$}0.3 & \cellcolor[HTML]{BCD7EB}28.2\,{\tiny$\pm$}1.6 & \cellcolor[HTML]{81BADB}44.8\,{\tiny$\pm$}3.2 & \cellcolor[HTML]{C3DAEE}26.0\,{\tiny$\pm$}0.9 \\
 & SE-16 & \cellcolor[HTML]{3383BE}\textcolor{white}{68.1\,{\tiny$\pm$}1.0} & \cellcolor[HTML]{92C4DE}40.4\,{\tiny$\pm$}0.4 & \cellcolor[HTML]{CEE0F2}21.0\,{\tiny$\pm$}1.1 & \cellcolor[HTML]{EEF5FC}4.5\,{\tiny$\pm$}1.2 & \cellcolor[HTML]{4D99CA}\textcolor{white}{59.4\,{\tiny$\pm$}1.4} & \cellcolor[HTML]{B3D3E8}30.9\,{\tiny$\pm$}2.1 & \cellcolor[HTML]{EAF3FB}6.6\,{\tiny$\pm$}1.8 \\
\addlinespace[2pt]
\multirow{3}{*}{\textbf{50--59}} & SE-15 & \cellcolor[HTML]{4191C6}\textcolor{white}{62.5\,{\tiny$\pm$}1.7} & \cellcolor[HTML]{C7DBEF}24.7\,{\tiny$\pm$}1.2 & \cellcolor[HTML]{D1E2F3}19.2\,{\tiny$\pm$}2.1 & \cellcolor[HTML]{DDEAF7}13.0\,{\tiny$\pm$}0.3 & \cellcolor[HTML]{95C5DF}39.5\,{\tiny$\pm$}3.0 & \cellcolor[HTML]{B4D3E9}30.6\,{\tiny$\pm$}4.2 & \cellcolor[HTML]{CEE0F2}20.8\,{\tiny$\pm$}1.0 \\
 & GE-15 & \cellcolor[HTML]{1663AA}\textcolor{white}{80.3\,{\tiny$\pm$}0.2} & \cellcolor[HTML]{C6DBEF}25.3\,{\tiny$\pm$}2.0 & \cellcolor[HTML]{9DCAE1}37.5\,{\tiny$\pm$}1.2 & \cellcolor[HTML]{D7E6F5}16.3\,{\tiny$\pm$}0.3 & \cellcolor[HTML]{B2D2E8}31.5\,{\tiny$\pm$}2.6 & \cellcolor[HTML]{79B5D9}46.7\,{\tiny$\pm$}1.6 & \cellcolor[HTML]{CFE1F2}20.3\,{\tiny$\pm$}0.4 \\
 & SE-16 & \cellcolor[HTML]{2070B4}\textcolor{white}{75.0\,{\tiny$\pm$}0.4} & \cellcolor[HTML]{97C6DF}39.4\,{\tiny$\pm$}1.5 & \cellcolor[HTML]{BCD7EB}28.3\,{\tiny$\pm$}0.8 & \cellcolor[HTML]{EEF5FC}4.4\,{\tiny$\pm$}0.5 & \cellcolor[HTML]{63A8D3}\textcolor{white}{52.5\,{\tiny$\pm$}2.2} & \cellcolor[HTML]{9DCAE1}37.8\,{\tiny$\pm$}1.3 & \cellcolor[HTML]{EBF3FB}5.9\,{\tiny$\pm$}0.6 \\
\addlinespace[2pt]
\multirow{3}{*}{\textbf{60--69}} & SE-15 & \cellcolor[HTML]{3A8AC2}\textcolor{white}{65.2\,{\tiny$\pm$}2.2} & \cellcolor[HTML]{B9D6EA}29.1\,{\tiny$\pm$}2.2 & \cellcolor[HTML]{CDE0F1}21.2\,{\tiny$\pm$}1.8 & \cellcolor[HTML]{E4EFF9}9.7\,{\tiny$\pm$}0.7 & \cellcolor[HTML]{81BADB}44.6\,{\tiny$\pm$}4.9 & \cellcolor[HTML]{AED1E7}32.5\,{\tiny$\pm$}4.0 & \cellcolor[HTML]{DAE8F6}14.8\,{\tiny$\pm$}1.6 \\
 & GE-15 & \cellcolor[HTML]{1562A9}\textcolor{white}{80.7\,{\tiny$\pm$}1.0} & \cellcolor[HTML]{B2D2E8}31.6\,{\tiny$\pm$}1.5 & \cellcolor[HTML]{A0CBE2}37.0\,{\tiny$\pm$}0.7 & \cellcolor[HTML]{E2EDF8}10.6\,{\tiny$\pm$}0.6 & \cellcolor[HTML]{97C6DF}39.1\,{\tiny$\pm$}2.3 & \cellcolor[HTML]{7CB7DA}45.9\,{\tiny$\pm$}1.5 & \cellcolor[HTML]{DDEAF7}13.2\,{\tiny$\pm$}0.9 \\
 & SE-16 & \cellcolor[HTML]{1966AD}\textcolor{white}{79.1\,{\tiny$\pm$}1.3} & \cellcolor[HTML]{92C4DE}40.5\,{\tiny$\pm$}2.2 & \cellcolor[HTML]{AFD1E7}32.4\,{\tiny$\pm$}0.5 & \cellcolor[HTML]{F1F7FD}3.2\,{\tiny$\pm$}1.0 & \cellcolor[HTML]{66ABD4}\textcolor{white}{51.3\,{\tiny$\pm$}3.6} & \cellcolor[HTML]{91C3DE}41.0\,{\tiny$\pm$}1.3 & \cellcolor[HTML]{EFF6FC}4.1\,{\tiny$\pm$}1.3 \\
\addlinespace[2pt]
\multirow{3}{*}{\textbf{70+}} & SE-15 & \cellcolor[HTML]{7AB6D9}46.5\,{\tiny$\pm$}3.7 & \cellcolor[HTML]{C8DCF0}24.2\,{\tiny$\pm$}3.1 & \cellcolor[HTML]{DCE9F6}13.7\,{\tiny$\pm$}1.1 & \cellcolor[HTML]{EEF5FC}4.6\,{\tiny$\pm$}0.2 & \cellcolor[HTML]{64A9D3}\textcolor{white}{52.0\,{\tiny$\pm$}10.9} & \cellcolor[HTML]{B8D5EA}29.5\,{\tiny$\pm$}4.8 & \cellcolor[HTML]{E3EEF9}10.0\,{\tiny$\pm$}1.3 \\
 & GE-15 & \cellcolor[HTML]{529DCC}\textcolor{white}{57.7\,{\tiny$\pm$}2.9} & \cellcolor[HTML]{C1D9ED}26.8\,{\tiny$\pm$}2.3 & \cellcolor[HTML]{C7DCEF}24.4\,{\tiny$\pm$}1.7 & \cellcolor[HTML]{EEF5FC}4.4\,{\tiny$\pm$}0.2 & \cellcolor[HTML]{7AB6D9}46.5\,{\tiny$\pm$}6.3 & \cellcolor[HTML]{8ABFDD}42.3\,{\tiny$\pm$}5.1 & \cellcolor[HTML]{E8F1FA}7.7\,{\tiny$\pm$}0.7 \\
 & SE-16 & \cellcolor[HTML]{4B98CA}\textcolor{white}{59.7\,{\tiny$\pm$}4.1} & \cellcolor[HTML]{B4D3E9}30.6\,{\tiny$\pm$}2.3 & \cellcolor[HTML]{C7DCEF}24.6\,{\tiny$\pm$}5.6 & \cellcolor[HTML]{F4F9FE}1.8\,{\tiny$\pm$}0.3 & \cellcolor[HTML]{66ABD4}\textcolor{white}{51.2\,{\tiny$\pm$}7.4} & \cellcolor[HTML]{8FC2DE}41.2\,{\tiny$\pm$}12.2 & \cellcolor[HTML]{F2F7FD}3.1\,{\tiny$\pm$}0.8 \\
\addlinespace[2pt]
\bottomrule
\end{tabular}
\end{table}

\begin{table}[!htb]
    \caption{The joint ethnicity-by-age crosstab, SE-16}
    \label{tab:joint}
    \centering{\footnotesize (point estimates without error bars are not citable)}\par\vspace{3pt}
    \footnotesize\setlength{\tabcolsep}{3.2pt}
    \begin{tabular}{lcccccc}
        \toprule
                      & 18--29                                                           & 30--39                                                           & 40--49                                                           & 50--59                                                           & 60--69                                                           & 70+                                                              \\
        \midrule
        \multicolumn{7}{l}{\textit{Turnout (\% of electorate)}}                                                                                                                                                                                                                                                                                                                                                                         \\[1pt]
        \quad Malay   & \cellcolor[HTML]{2171B5}\textcolor{white}{74.6\,{\tiny$\pm$}2.5} & \cellcolor[HTML]{2575B7}\textcolor{white}{73.1\,{\tiny$\pm$}2.5} & \cellcolor[HTML]{2373B6}\textcolor{white}{74.0\,{\tiny$\pm$}2.5} & \cellcolor[HTML]{1460A8}\textcolor{white}{81.4\,{\tiny$\pm$}2.5} & \cellcolor[HTML]{105BA4}\textcolor{white}{83.5\,{\tiny$\pm$}2.5} & \cellcolor[HTML]{3282BE}\textcolor{white}{68.7\,{\tiny$\pm$}4.1} \\
        \quad Chinese & \cellcolor[HTML]{61A7D2}\textcolor{white}{52.9\,{\tiny$\pm$}2.2} & \cellcolor[HTML]{5CA4D0}\textcolor{white}{54.6\,{\tiny$\pm$}2.2} & \cellcolor[HTML]{4896C8}\textcolor{white}{60.7\,{\tiny$\pm$}2.2} & \cellcolor[HTML]{3181BD}\textcolor{white}{68.8\,{\tiny$\pm$}2.2} & \cellcolor[HTML]{1D6CB1}\textcolor{white}{76.6\,{\tiny$\pm$}2.2} & \cellcolor[HTML]{56A0CE}\textcolor{white}{56.3\,{\tiny$\pm$}4.1} \\
        \quad Indian  & 49.5                                                             & 47.5                                                             & 63.0                                                             & 65.3                                                             & 73.2                                                             & 29.1                                                             \\
        \quad Other   & 60.6                                                             & 75.6                                                             & 57.7                                                             & 84.8                                                             & 33.8                                                             & 49.8                                                             \\
        \addlinespace[3pt]
        \multicolumn{7}{l}{\textit{BN (\% of electorate)}}                                                                                                                                                                                                                                                                                                                                                                              \\[1pt]
        \quad Malay   & \cellcolor[HTML]{3C8CC3}\textcolor{white}{64.8\,{\tiny$\pm$}1.0} & \cellcolor[HTML]{4191C6}\textcolor{white}{62.6\,{\tiny$\pm$}1.0} & \cellcolor[HTML]{4292C6}\textcolor{white}{62.1\,{\tiny$\pm$}1.0} & \cellcolor[HTML]{3888C1}\textcolor{white}{66.3\,{\tiny$\pm$}1.5} & \cellcolor[HTML]{2F7FBC}\textcolor{white}{69.9\,{\tiny$\pm$}2.2} & \cellcolor[HTML]{4D99CA}\textcolor{white}{59.1\,{\tiny$\pm$}2.3} \\
        \quad Chinese & \cellcolor[HTML]{F2F7FD}2.8\,{\tiny$\pm$}0.9                     & \cellcolor[HTML]{EEF5FC}4.5\,{\tiny$\pm$}0.5                     & \cellcolor[HTML]{E9F2FA}7.2\,{\tiny$\pm$}0.4                     & \cellcolor[HTML]{E5EFF9}9.1\,{\tiny$\pm$}1.5                     & \cellcolor[HTML]{E3EEF9}10.0\,{\tiny$\pm$}2.2                    & \cellcolor[HTML]{E3EEF9}9.9\,{\tiny$\pm$}2.3                     \\
        \quad Indian  & 24.2                                                             & 22.9                                                             & 34.4                                                             & 30.5                                                             & 34.4                                                             & 13.1                                                             \\
        \quad Other   & 48.7                                                             & 64.0                                                             & 38.1                                                             & 45.0                                                             & 17.5                                                             & 0.7                                                              \\
        \addlinespace[3pt]
        \multicolumn{7}{l}{\textit{PH (\% of electorate)}}                                                                                                                                                                                                                                                                                                                                                                              \\[1pt]
        \quad Malay   & \cellcolor[HTML]{F1F7FD}3.2\,{\tiny$\pm$}0.4                     & \cellcolor[HTML]{F1F7FD}3.3\,{\tiny$\pm$}0.4                     & \cellcolor[HTML]{F2F7FD}3.1\,{\tiny$\pm$}1.1                     & \cellcolor[HTML]{EEF5FC}4.8\,{\tiny$\pm$}0.8                     & \cellcolor[HTML]{EDF4FC}5.3\,{\tiny$\pm$}0.5                     & \cellcolor[HTML]{F1F7FD}3.2\,{\tiny$\pm$}5.6                     \\
        \quad Chinese & \cellcolor[HTML]{72B2D8}48.3\,{\tiny$\pm$}0.3                    & \cellcolor[HTML]{71B1D7}48.6\,{\tiny$\pm$}0.3                    & \cellcolor[HTML]{65AAD4}\textcolor{white}{51.6\,{\tiny$\pm$}1.1} & \cellcolor[HTML]{529DCC}\textcolor{white}{57.7\,{\tiny$\pm$}0.8} & \cellcolor[HTML]{3C8CC3}\textcolor{white}{64.6\,{\tiny$\pm$}0.5} & \cellcolor[HTML]{84BCDB}44.0\,{\tiny$\pm$}5.6                    \\
        \quad Indian  & 15.4                                                             & 15.7                                                             & 18.4                                                             & 21.1                                                             & 24.1                                                             & 7.8                                                              \\
        \quad Other   & 5.9                                                              & 4.6                                                              & 6.4                                                              & 22.6                                                             & 2.8                                                              & 26.4                                                             \\
        \addlinespace[3pt]
        \multicolumn{7}{l}{\textit{PN (\% of electorate)}}                                                                                                                                                                                                                                                                                                                                                                              \\[1pt]
        \quad Malay   & \cellcolor[HTML]{ECF4FB}5.7\,{\tiny$\pm$}1.0                     & \cellcolor[HTML]{EAF3FB}6.5\,{\tiny$\pm$}0.7                     & \cellcolor[HTML]{E7F1FA}8.2\,{\tiny$\pm$}1.2                     & \cellcolor[HTML]{E5EFF9}9.1\,{\tiny$\pm$}0.7                     & \cellcolor[HTML]{EAF2FB}6.7\,{\tiny$\pm$}1.0                     & \cellcolor[HTML]{EFF6FC}4.1\,{\tiny$\pm$}0.7                     \\
        \quad Chinese & \cellcolor[HTML]{F7FBFF}0.0\,{\tiny$\pm$}1.0                     & \cellcolor[HTML]{F7FBFF}0.0\,{\tiny$\pm$}0.6                     & \cellcolor[HTML]{F7FBFF}0.0\,{\tiny$\pm$}1.2                     & \cellcolor[HTML]{F7FBFF}0.0\,{\tiny$\pm$}0.5                     & \cellcolor[HTML]{F7FBFF}0.0\,{\tiny$\pm$}1.0                     & \cellcolor[HTML]{F7FBFF}0.3\,{\tiny$\pm$}0.3                     \\
        \quad Indian  & 0.0                                                              & 0.0                                                              & 0.0                                                              & 0.0                                                              & 0.0                                                              & 0.0                                                              \\
        \quad Other   & 0.0                                                              & 0.0                                                              & 0.0                                                              & 0.1                                                              & 0.1                                                              & 0.2                                                              \\
        \addlinespace[3pt]
        \bottomrule
    \end{tabular}
\end{table}

\subsection{Participation}
\label{sec:findings-turnout}

\begin{figure}[!htb]
	\centering
	\caption{Turnout against ethnic composition, GE-15 vs SE-16}
	\includegraphics[width=\textwidth]{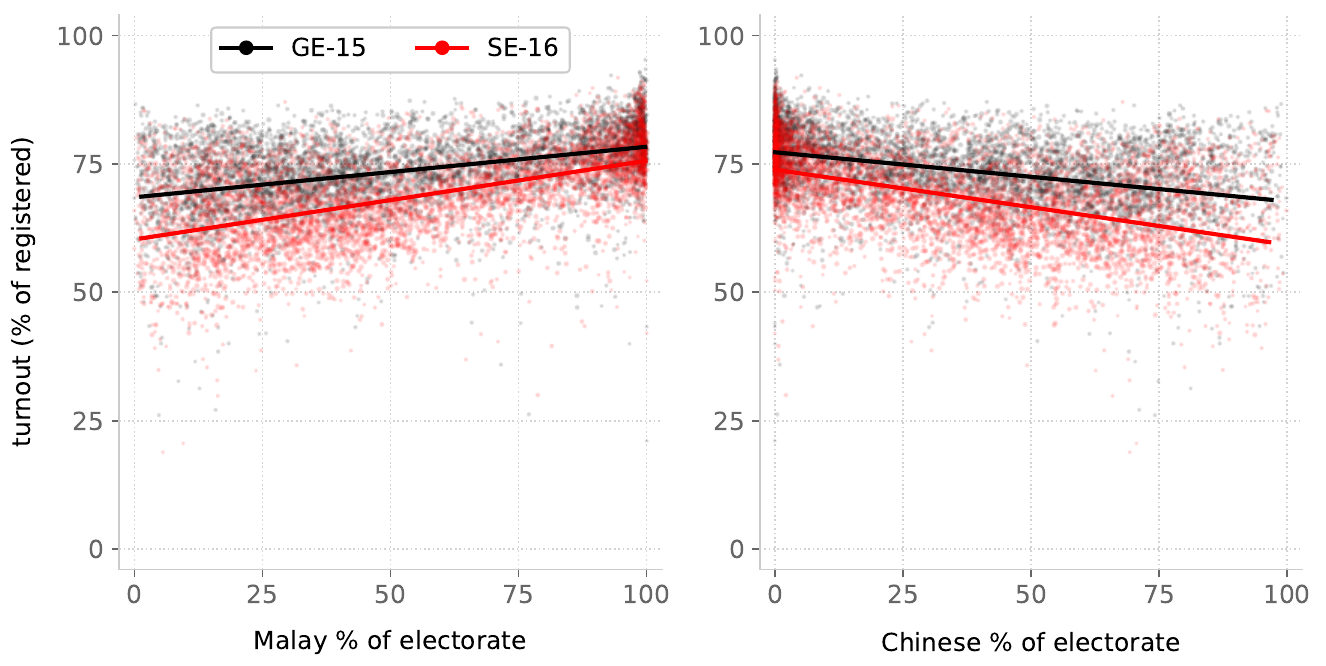}
	\label{fig:turnout}
\end{figure}

Turnout varies more across the three elections than any coalition rate. Chinese turnout rose from 42.8\% of the Chinese electorate at SE-15 to 69.0\% at GE-15 before falling to 61.0\% at SE-16; Malay turnout moved from 63.1\% to 78.3\% to 75.5\%. The ethnic gap therefore narrowed from twenty points to nine, then widened to fourteen. Figure~\ref{fig:turnout} shows the final step directly: turnout fell everywhere between GE-15 and SE-16, but roughly twice as steeply in Chinese-dominated salurans as in Malay-dominated ones. The exceptionally low SE-15 Chinese rate plausibly also reflects the then still-restricted Singapore border (Section~\ref{sec:context}).

\begin{figure}[!htb]
	\centering
	\caption{Estimated turnout by ethnicity}
	\includegraphics[width=\textwidth]{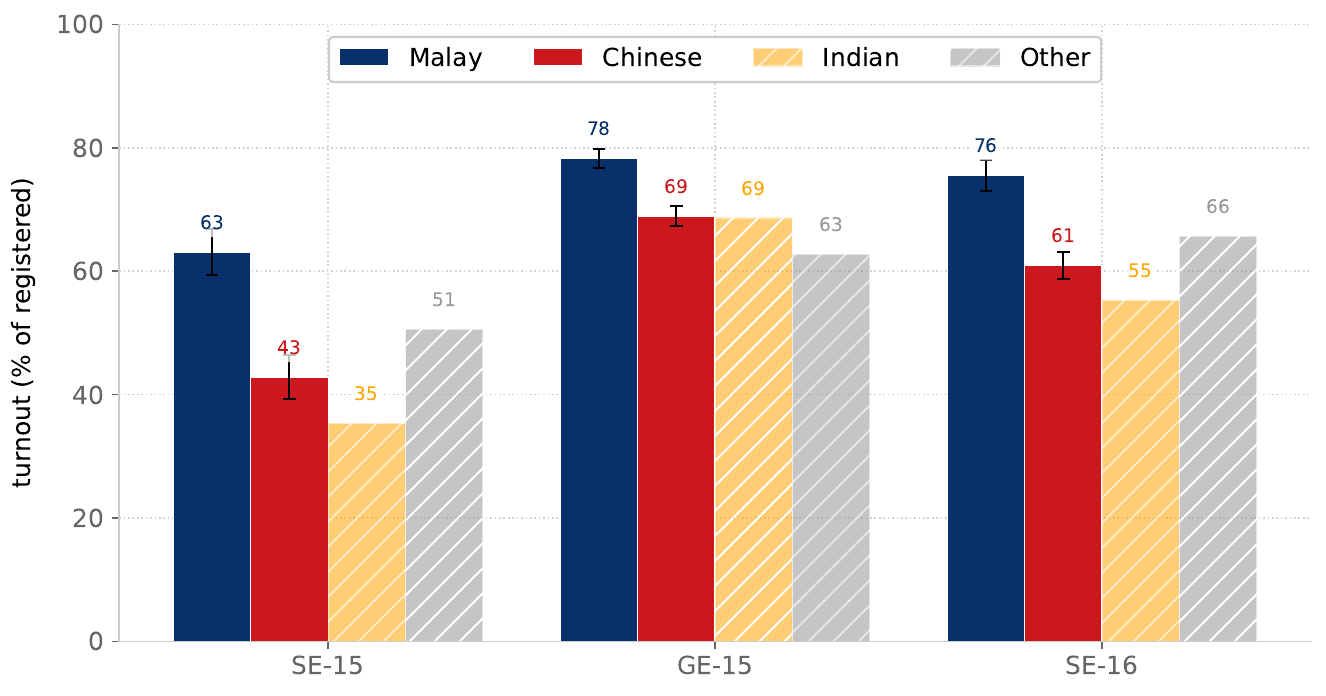}
	\label{fig:turnout-eth}
\end{figure}

\begin{figure}[!htb]
	\centering
	\caption{Estimated turnout by age band}
	\includegraphics[width=\textwidth]{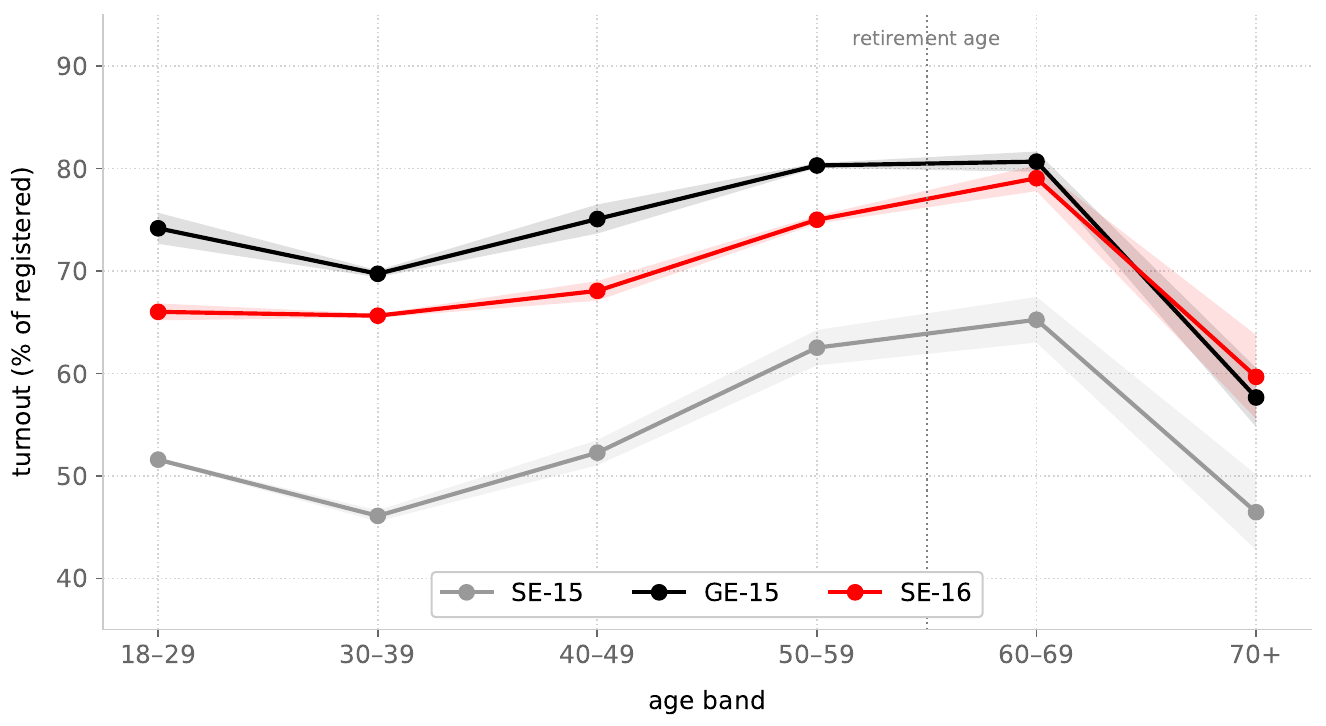}
	\label{fig:turnout-age}
\end{figure}

Across age bands, the most interesting feature sits at Malaysia's retirement age. At SE-16 turnout rises from a 65.6\% trough among the 30--39s to 79.1\% among the 60--69s---the decade \emph{after} retirement---then falls to 59.7\% past 70, a peak-to-trough spread of 13.4 points against a combined validation error of 1.4 (Figure~\ref{fig:turnout-age}). Retirement does not demobilise: the 60--69s out-vote the 50--59s by 4.1 points, about two and a half times that gap's combined error, and the same ordering holds at all three elections.

The trough is likewise not the youngest band but the thirties, whose turnout at both 2022 elections was the lowest of any band under 70---46.1\% at SE-15, below even the 51.6\% of the supposedly apathetic 18--29s. This life-cycle profile---a dip in the young-family decade, a peak in the retirement decade, and a cliff at 70---survives within both major ethnic groups rather than reflecting composition: within Malay voters, SE-16 turnout runs 74.6, 73.1, 74.0, 81.4, 83.5 and 68.7\% across the bands, and the Chinese profile has the same shape at a lower level with a steeper gradient (Table~\ref{tab:joint}).

The thirties are therefore a more specific mobilisation target than `the youth', while growth in the 70+ share will mechanically weigh on aggregate turnout, since the collapse past 70 runs 19 points below the peak. These data cannot distinguish practical barriers from declining interest as the explanation.

\subsection{The Malay reversal}
\label{sec:findings-malay}

\begin{figure}[!htb]
	\centering
	\caption{Coalition vote share against Malay composition, GE-15 vs SE-16}
	\includegraphics[width=\textwidth]{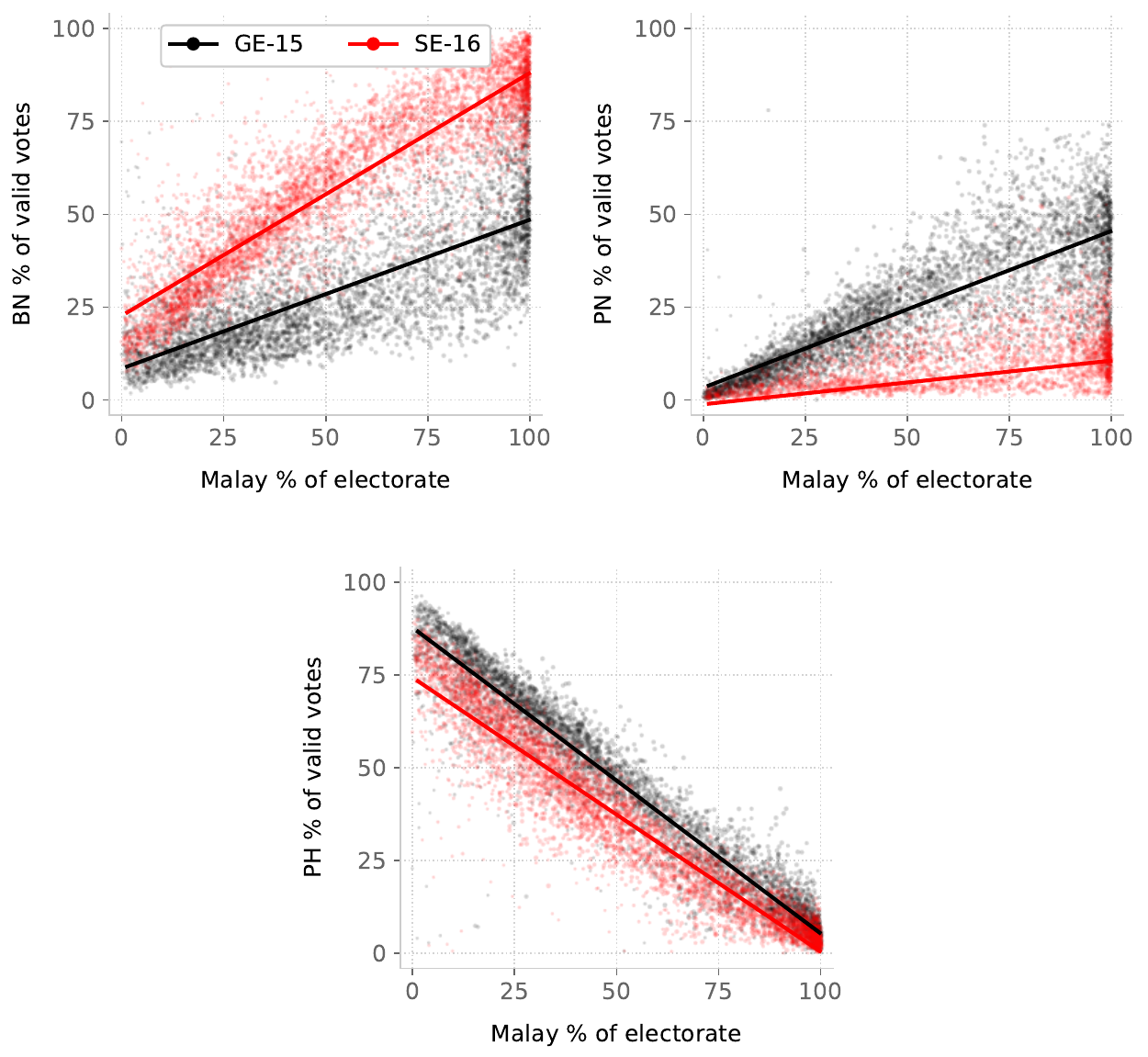}
	\label{fig:scatter-malay}
\end{figure}

Malay BN support was flat across the two 2022 elections---36.0\% and 36.5\% of the Malay electorate---and then rose to 64.1\% at SE-16, a move of $+27.6$ points against a combined validation error of 7.1. Malay PN support fell from 34.5\% to 6.8\% over the same step, a move of $-27.7$ against 3.2. The movements are equal and opposite, and no other Malay flow is large enough to account for either: PH stayed in single digits throughout, and Malay turnout moved less than 3 points, within its validation error. On the ballot scale, Malays gave BN 84.9\% of their ballots at SE-16 against 46.7\% at GE-15, a swing of roughly 430,000 ballots against a statewide BN gain of 513,809. Figure~\ref{fig:scatter-malay} shows the same rotation in the raw saluran relationship: BN's share climbs steeply with Malay composition---from about 48\% of the valid vote in near-fully-Malay salurans to nearly 90\%---while the PN line collapses towards the axis and the PH line barely moves.

PN's support among non-Malays is essentially zero at every election, so PN's Johor story is entirely a Malay story. Within it, the crosstab reveals a structure invisible in the marginals: the 2022 wave was steeply age-graded. Table~\ref{tab:malay} shows it on the scale where it is most legible, each band's support as a share of its own ballots. At GE-15 PN won 52.7\% of ballots among 18--29-year-old Malays, against 24.7\% among the 60--69s and 14.1\% among the over-70s, while BN's shares ran the other way, from 34.7\% among the youngest to 78.4\% among the oldest. By SE-16 every band gave BN 81--87\% and PN 6--11\%. The reversal was therefore largest exactly where the wave had been strongest: the statewide movement is the sum of age-specific reversals, not a uniform shift.

\begin{table}[!htb]
\caption{The Malay reversal by age band, across the three elections}
\label{tab:malay}
\centering
\footnotesize\setlength{\tabcolsep}{3.2pt}
\begin{tabular}{lcccccc}
\toprule
 & 18--29 & 30--39 & 40--49 & 50--59 & 60--69 & 70+ \\
\midrule
\multicolumn{7}{l}{\textit{Turnout (\% of Malay electorate)}}\\[1pt]
\quad SE-15 & \cellcolor[HTML]{4493C7}\textcolor{white}{61.8\,{\tiny$\pm$}3.7} & \cellcolor[HTML]{58A1CF}\textcolor{white}{55.9\,{\tiny$\pm$}3.7} & \cellcolor[HTML]{3F8FC5}\textcolor{white}{63.3\,{\tiny$\pm$}3.7} & \cellcolor[HTML]{2676B8}\textcolor{white}{72.8\,{\tiny$\pm$}3.7} & \cellcolor[HTML]{2474B7}\textcolor{white}{73.5\,{\tiny$\pm$}3.7} & \cellcolor[HTML]{4F9BCB}\textcolor{white}{58.3\,{\tiny$\pm$}3.7} \\
\quad GE-15 & \cellcolor[HTML]{1764AB}\textcolor{white}{79.8\,{\tiny$\pm$}1.6} & \cellcolor[HTML]{2474B7}\textcolor{white}{73.8\,{\tiny$\pm$}1.6} & \cellcolor[HTML]{1A68AE}\textcolor{white}{78.5\,{\tiny$\pm$}1.6} & \cellcolor[HTML]{0E59A2}\textcolor{white}{84.3\,{\tiny$\pm$}1.6} & \cellcolor[HTML]{0E58A2}\textcolor{white}{84.6\,{\tiny$\pm$}1.6} & \cellcolor[HTML]{3E8EC4}\textcolor{white}{63.9\,{\tiny$\pm$}2.9} \\
\quad SE-16 & \cellcolor[HTML]{2171B5}\textcolor{white}{74.6\,{\tiny$\pm$}2.5} & \cellcolor[HTML]{2575B7}\textcolor{white}{73.1\,{\tiny$\pm$}2.5} & \cellcolor[HTML]{2373B6}\textcolor{white}{74.0\,{\tiny$\pm$}2.5} & \cellcolor[HTML]{1460A8}\textcolor{white}{81.4\,{\tiny$\pm$}2.5} & \cellcolor[HTML]{105BA4}\textcolor{white}{83.5\,{\tiny$\pm$}2.5} & \cellcolor[HTML]{3282BE}\textcolor{white}{68.7\,{\tiny$\pm$}4.1} \\
\addlinespace[3pt]
\multicolumn{7}{l}{\textit{BN (\% of Malay ballots)}}\\[1pt]
\quad SE-15 & \cellcolor[HTML]{63A8D3}\textcolor{white}{52.7\,{\tiny$\pm$}11.9} & \cellcolor[HTML]{5FA6D1}\textcolor{white}{53.7\,{\tiny$\pm$}13.3} & \cellcolor[HTML]{5FA6D1}\textcolor{white}{53.7\,{\tiny$\pm$}11.7} & \cellcolor[HTML]{549FCD}\textcolor{white}{56.7\,{\tiny$\pm$}10.3} & \cellcolor[HTML]{2F7FBC}\textcolor{white}{69.7\,{\tiny$\pm$}10.9} & \cellcolor[HTML]{1E6DB2}\textcolor{white}{76.4\,{\tiny$\pm$}14.1} \\
\quad GE-15 & \cellcolor[HTML]{A8CEE4}34.7\,{\tiny$\pm$}8.3 & \cellcolor[HTML]{89BEDC}42.9\,{\tiny$\pm$}9.1 & \cellcolor[HTML]{7AB6D9}46.4\,{\tiny$\pm$}8.7 & \cellcolor[HTML]{65AAD4}\textcolor{white}{51.7\,{\tiny$\pm$}8.2} & \cellcolor[HTML]{3383BE}\textcolor{white}{68.3\,{\tiny$\pm$}8.5} & \cellcolor[HTML]{1A68AE}\textcolor{white}{78.4\,{\tiny$\pm$}13.0} \\
\quad SE-16 & \cellcolor[HTML]{09529D}\textcolor{white}{86.8\,{\tiny$\pm$}4.2} & \cellcolor[HTML]{0B559F}\textcolor{white}{85.7\,{\tiny$\pm$}4.3} & \cellcolor[HTML]{0E59A2}\textcolor{white}{84.0\,{\tiny$\pm$}4.2} & \cellcolor[HTML]{1460A8}\textcolor{white}{81.4\,{\tiny$\pm$}4.3} & \cellcolor[HTML]{0F5AA3}\textcolor{white}{83.7\,{\tiny$\pm$}5.2} & \cellcolor[HTML]{0A549E}\textcolor{white}{86.1\,{\tiny$\pm$}8.5} \\
\addlinespace[3pt]
\multicolumn{7}{l}{\textit{PN (\% of Malay ballots)}}\\[1pt]
\quad SE-15 & \cellcolor[HTML]{9CC9E1}38.0\,{\tiny$\pm$}3.7 & \cellcolor[HTML]{97C6DF}39.3\,{\tiny$\pm$}4.2 & \cellcolor[HTML]{99C7E0}38.8\,{\tiny$\pm$}3.7 & \cellcolor[HTML]{A9CFE5}34.0\,{\tiny$\pm$}3.0 & \cellcolor[HTML]{C8DCF0}24.2\,{\tiny$\pm$}2.5 & \cellcolor[HTML]{D9E8F5}15.1\,{\tiny$\pm$}2.5 \\
\quad GE-15 & \cellcolor[HTML]{61A7D2}\textcolor{white}{52.7\,{\tiny$\pm$}4.1} & \cellcolor[HTML]{6DAFD7}49.4\,{\tiny$\pm$}4.4 & \cellcolor[HTML]{79B5D9}46.6\,{\tiny$\pm$}4.1 & \cellcolor[HTML]{9AC8E0}38.6\,{\tiny$\pm$}3.6 & \cellcolor[HTML]{C7DBEF}24.7\,{\tiny$\pm$}3.4 & \cellcolor[HTML]{DBE9F6}14.1\,{\tiny$\pm$}4.5 \\
\quad SE-16 & \cellcolor[HTML]{E8F1FA}7.6\,{\tiny$\pm$}1.6 & \cellcolor[HTML]{E6F0F9}8.9\,{\tiny$\pm$}1.2 & \cellcolor[HTML]{E1EDF8}11.0\,{\tiny$\pm$}2.0 & \cellcolor[HTML]{E1EDF8}11.1\,{\tiny$\pm$}1.2 & \cellcolor[HTML]{E7F1FA}8.0\,{\tiny$\pm$}1.4 & \cellcolor[HTML]{EBF3FB}5.9\,{\tiny$\pm$}1.4 \\
\addlinespace[3pt]
\multicolumn{7}{l}{\textit{PH (\% of Malay ballots)}}\\[1pt]
\quad SE-15 & \cellcolor[HTML]{EEF5FC}4.9\,{\tiny$\pm$}1.5 & \cellcolor[HTML]{F0F6FD}3.8\,{\tiny$\pm$}1.6 & \cellcolor[HTML]{F0F6FD}3.7\,{\tiny$\pm$}2.5 & \cellcolor[HTML]{EEF5FC}4.6\,{\tiny$\pm$}3.1 & \cellcolor[HTML]{F2F7FD}2.9\,{\tiny$\pm$}2.5 & \cellcolor[HTML]{F2F7FD}3.0\,{\tiny$\pm$}2.1 \\
\quad GE-15 & \cellcolor[HTML]{E0ECF8}11.7\,{\tiny$\pm$}3.9 & \cellcolor[HTML]{EAF2FB}6.7\,{\tiny$\pm$}2.8 & \cellcolor[HTML]{EBF3FB}6.1\,{\tiny$\pm$}2.6 & \cellcolor[HTML]{E7F1FA}8.0\,{\tiny$\pm$}2.5 & \cellcolor[HTML]{EDF4FC}5.3\,{\tiny$\pm$}2.4 & \cellcolor[HTML]{EFF6FC}4.0\,{\tiny$\pm$}3.2 \\
\quad SE-16 & \cellcolor[HTML]{EFF6FC}4.3\,{\tiny$\pm$}0.7 & \cellcolor[HTML]{EEF5FC}4.5\,{\tiny$\pm$}0.7 & \cellcolor[HTML]{EFF6FC}4.2\,{\tiny$\pm$}1.6 & \cellcolor[HTML]{EBF3FB}5.9\,{\tiny$\pm$}1.2 & \cellcolor[HTML]{EAF3FB}6.3\,{\tiny$\pm$}0.8 & \cellcolor[HTML]{EEF5FC}4.6\,{\tiny$\pm$}8.4 \\
\addlinespace[3pt]
\bottomrule
\end{tabular}
\end{table}

\subsubsection*{Did PN simply hand its voters to BN?}

PN's stand-down in 23 seats creates the main rival explanation: BN may have gained through coalition coordination rather than changed voter preference. Three checks reject that account, all comparing GE-15 with SE-16 on fixed territory, since Johor's 949 polling districts are identical at the two elections.

\begin{figure}[!htb]
	\centering
	\caption{The GE-15 to SE-16 swing, by whether PN contested the seat}
	\includegraphics[width=\textwidth]{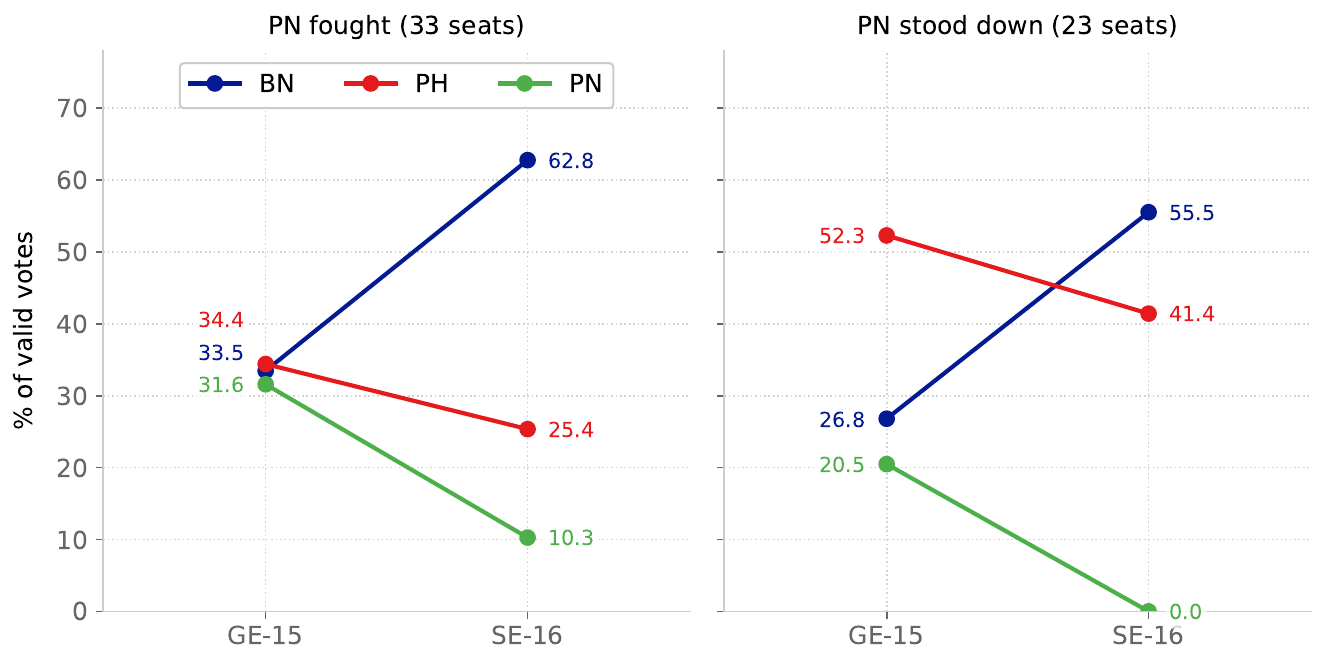}
	\label{fig:standdown}
\end{figure}

First, if the stand-down were doing the work, BN should have gained more where PN withdrew. It did not: BN gained 29.3 points of the valid vote in the 33 seats PN fought and 28.7 in the 23 it left, a difference of 0.6 points in the wrong direction (Figure~\ref{fig:standdown}).

Second, vanished PN votes account for almost the same share of BN's gain on both sides of the divide: 78.1\% in the stand-down seats, against 77.5\% in the contested seats, where PN's collapse is directly observed.

Third, and decisively: within the 33 seats PN did contest, take the 198 polling districts more than 90\% Malay, where the ecological arithmetic nearly pins the answer down on its own. Measured against the electorate on fixed territory, BN rose from 39.1\% to 61.0\% and PN fell from 32.1\% to 10.4\%---a gain of 21.8 points against a loss of 21.7, a ratio of 1.01, holding at 0.99--1.02 at purity thresholds of 80 or 95\%---while PH moved 2.4 points and turnout 2.2.

The one-for-one relationship therefore appears in near-model-free form exactly where the stand-down cannot contribute, and the 23 vacated seats hold only 55 of the 253 heavily Malay districts, so they are not carrying the finding.

\subsubsection*{What the pattern rules out, and what it suggests}

The pattern is inconsistent with a mobilisation account, because Malay turnout did not rise, and the measurable non-Malay contributions were small. It therefore identifies net reversion within the Malay electorate. The natural interpretation is a return from PN to BN, although individual transitions and offsetting churn remain unobserved.

Three contextual mechanisms are consistent with the reversal: the change from a scandal-damaged UMNO in 2022 to a popular BN state administration amid economic growth, which is the circumstance under which an economic-grievance protest rather than a durable ethno-religious realignment should recede \cite{washida2023}; PN's visible organisational disintegration in the months before SE-16; and the greater weight of state machinery in a state election, where UMNO Johor's organisation has no PN counterpart \cite{ostwaldoliver2023}. These are interpretations, not identified causal effects.

The age structure weighs against a durable cohort realignment: the 2022 gradient disappeared rather than ageing forward with the cohort, and by SE-16 the 18--29s were indistinguishable from their elders, residual PN support peaking instead at 11.1\% of ballots in the 50--59 band (Table~\ref{tab:malay}). The wave was generational in its arrival and general in its retreat.

The arrangement Johor previewed has since been extended and made explicit: heading into the Negeri Sembilan state election (to be held on 1 Aug 2026), BN and PN have divided all 36 seats between them. Two Johor observations carry over: first, BN's gain was no larger where PN withdrew, so standing down added little that BN was not achieving anyway; second, PN's position is weaker than its headline suggests, because the seats it kept were the most favourable to it and it still lost more than two thirds of its vote there. These are valuable baseline findings, because the Negeri Sembilan election---where there is no constituency with both a BN and PN candidate contesting---will not offer the same scope for study.

Finally, Johor is not Malaysia: it is UMNO's heartland and PN's weakest peninsular flank. Accordingly, although our findings show that the PN-to-BN flow can happen fast and at near-total levels, we do not expect it to be repeatable at the same scale elsewhere.

\subsection{Chinese demobilisation}
\label{sec:findings-chinese}

\begin{figure}[!htb]
	\centering
	\caption{Coalition vote share against Chinese composition, GE-15 vs SE-16}
	\includegraphics[width=\textwidth]{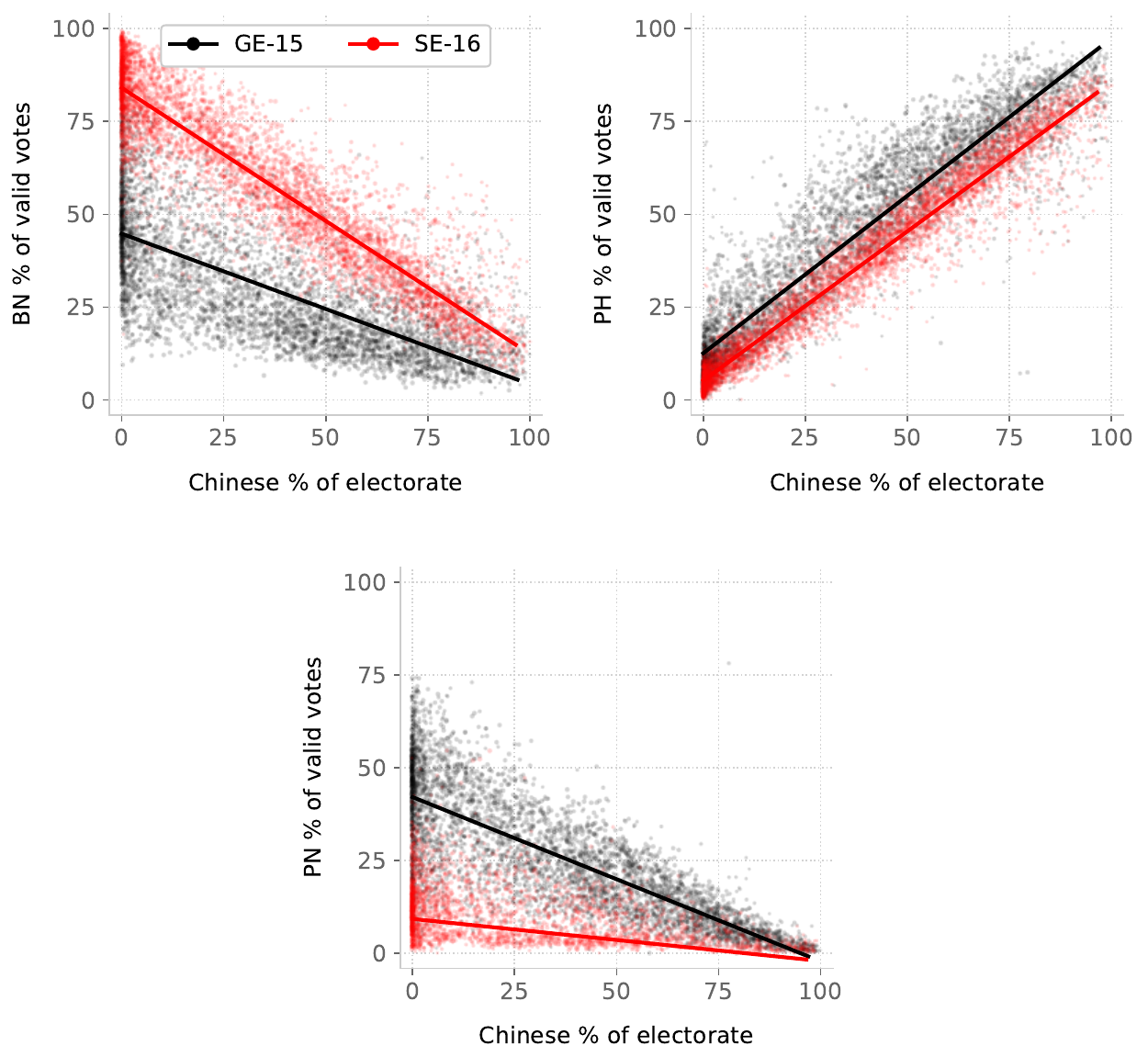}
	\label{fig:scatter-chinese}
\end{figure}

PH support fell from 63.1\% to 52.2\% of the Chinese electorate between GE-15 and SE-16, a move of 10.9 points against a combined error of 0.6. Chinese turnout fell 8.0 points against 3.8, while PH's share among Chinese voters fell 5.9 points, from 91.5\% to 85.6\%, against a combined error of 6.1. Roughly two thirds of the electorate-scale decline therefore came through abstention, and the remaining third rests on a change we cannot distinguish from zero. BN's share of Chinese ballots rose from 6.6\% to 11.2\% on the same weak footing, and Chinese BN support across the three elections---4.5, 4.6 and 6.8\% of the Chinese electorate---is the quietest series in Table~\ref{tab:results}: the 2.2-point movement at SE-16 is real, at more than four times its combined error, but an order of magnitude smaller than the Malay swing. Figure~\ref{fig:scatter-chinese} shows the shift in the raw data: the PH line against Chinese composition moves down roughly in parallel, while the BN line rises modestly from a low base.

This places Johor's Chinese electorate in a pattern familiar from the comparative literature on differential abstention: when a community's preferred party faces an unlosable or unwinnable election, or its alternatives have converged, participation absorbs the shock that vote-switching would otherwise register. All three conditions held at SE-16: the outcome was not in doubt, PH's federal alliance with BN meant a DAP voter was indirectly supporting a government containing UMNO, and PN was visibly moribund in the state. Contemporaneous commentary read the result the same way---Chinese voters staying home to punish PH rather than crossing over \cite{rsis2026}. What the estimates add is that only one of that account's two components is established: `predominantly demobilisation' is the strongest claim these data support.

The SE-15 comparison points the same way, with one caveat. Chinese turnout at SE-15 sat 20 points below Malay turnout, the widest gap in the series---but SE-15 was held three weeks before the Singapore border reopened to quarantine-free travel, and Chinese Johoreans are overrepresented in the cross-border workforce, so part of that gap is plausibly a travel barrier rather than a choice. The cleaner evidence that demobilisation is a standing response to low-stakes elections rather than a 2026 novelty is SE-16 itself, where no restriction was in place and the gap was still fourteen points. The persistence of the gap carries a competitive implication: in a seat-based system a 14-point ethnic turnout differential can convert PH's efficient urban vote into fewer seats, which is exactly what happened---PH won 8 of 56 seats on a third of the valid vote. Whether Chinese participation rebounds at the next general election, as it did between SE-15 and GE-15 with a 26-point jump in eight months, is arguably the most consequential open parameter for GE-16 in mixed seats.

\subsection{Age: Mostly composition, not behaviour}
\label{sec:findings-age}

\begin{figure}[!htb]
	\centering
	\caption{Coalition vote share against the saluran's median age, GE-15 vs SE-16}
	\includegraphics[width=\textwidth]{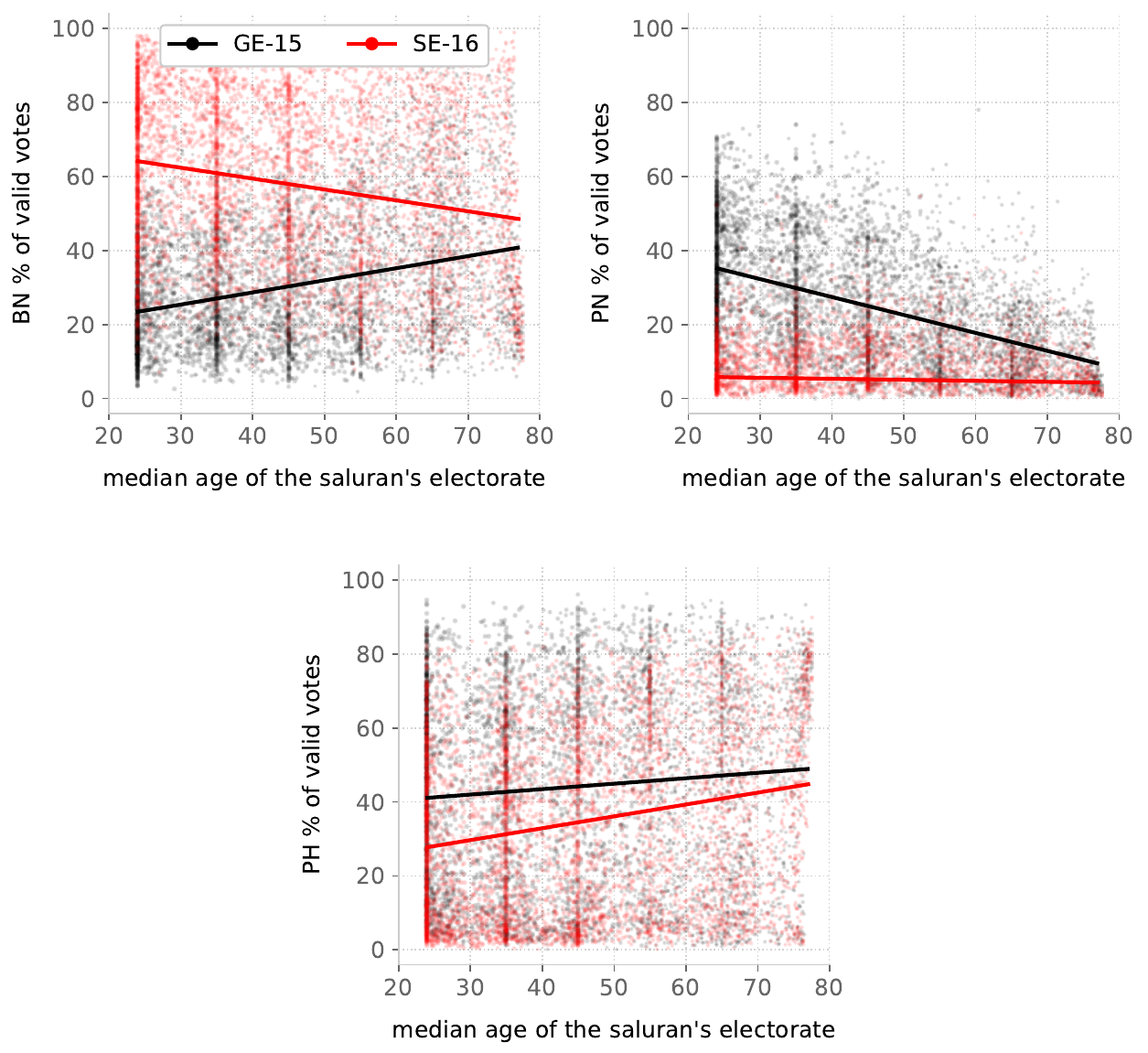}
	\label{fig:age-scatter}
\end{figure}

Marginally, BN support appears strongest among younger voters at SE-16. Figure~\ref{fig:age-scatter} shows support varying steeply and smoothly with a saluran's median age, and the marginals agree: BN's age profile inverts between the elections, rising with age at GE-15 (17.8\% of the 18--29 electorate to 31.6\% of the 60--69s) and falling with age at SE-16 (42.9\% among the youngest down to 30.6\% among the over-70s). Both the reversal and the 2026 gradient survive their errors, inviting the conclusion that BN's recovery was led by the young.

Within ethnicity, that gradient largely disappears. Malay support for each coalition is nearly flat across age (Table~\ref{tab:malay}: BN between 81 and 87\% of each band's ballots, PN 6--11\%, PH 4--6\%), while Chinese variation operates mainly through the turnout life-cycle of Section~\ref{sec:findings-turnout} rather than through vote choice. The marginal pattern arises because younger cohorts are more Malay and older cohorts more Chinese (Figure~\ref{fig:agecomp}). GE-15 differs: young Malays genuinely preferred PN, so composition and within-Malay behaviour reinforced one another. By SE-16 the behavioural gradient had collapsed, leaving only the unchanged compositional gradient---so what inverted between the elections was not behaviour twisting from one gradient to its opposite. The crosstab's authority for these statements comes from the holdout, since age structure is exactly what the joint model was shown to recover.

This clarifies a live Malaysian argument. Because the enfranchisement reforms made the youngest cohort the most Malay, any pro-Malay-party swing appears youth-led in marginal data, and debates over whether Undi18 fuelled the green wave have turned on exactly such comparisons. The young Malay tilt to PN at GE-15 was real and large, but it appeared only after SE-15---polling-district analyses of the newly enfranchised there found no distinctive behaviour \cite{zhanghutchinson2022}---and had disappeared by SE-16. The apparent 2026 youth tilt to BN is instead what Johor's age-by-ethnicity composition predicts.

\section{Broader implications for EI applications}
\label{sec:implications}

Johor's electoral geography is unusually favourable to ecological inference, and its sorting rule is an administrative stroke of luck, albeit one which is likely to persist as long as the EC does not change its administrative practices. However, we identify three lessons from this exercise which are broadly applicable to EI applications.

First, staying home is a choice, and should be modelled as one. Most EI applications condition on having voted, estimating how each group divided the ballots it cast---which may end up removing a large source of inter-election variation. Making the electorate the denominator instead means a group's support rates sum to its turnout, and its turnout and abstention to one, keeping mobilisation and persuasion inside a single frame. Johor's Chinese electorate shows what this buys: PH's support fell 10.9 points of that electorate between GE-15 and SE-16, roughly two-thirds of it through lower turnout. Conditioning on voters would have shown only the remaining third---a 5.9-point fall in PH's ballot share that we cannot distinguish from zero once our error bars are included---missing the finding while reporting a smaller one confidently. The distinction is not cosmetic, because the two channels imply different prospects for recovery: a party that has lost voters to abstention faces a different task from one that has lost them to a rival.

Second, identification is won through richer rows, not richer or more sophisticated likelihoods. Age-by-vote tabulations are among the most-quoted numbers in electoral commentary anywhere, and wherever age composition differs along a cleavage which also affects behaviour (such as ethnicity), there is a clear omitted variable problem, and the marginal age gradient is confounded by construction \cite{achenshively1995,chen2026}---in our paper, the magnitude of the bias is enough to reverse the sign of some findings. The fix is not a better estimator but a finer cross-classification: our deliberately plain specification matched a hierarchical Bayesian R$\times$C model cell-for-cell on the held-out streams, while redefining a row from 10 marginal groups to 24 joint cells cut the age-cell miss from 4.93 points to 0.98. Where the row definition omits a jointly observed confounder, a finer cross-classification should take priority over a more sophisticated likelihood---and whether one is available is a question about data, not about methodology.

Third, uncertainty should be measured where it can be and left unquantified where it cannot. The trap is to reach for a classical standard error, which asks how far an estimate would move in a different sample, with the property that the standard error shrinks as the sample grows; an EI estimate's problem is not that it was sampled (in fact, in Malaysia's case, the entire universe of data is observed) but that it is unidentified given the data that are available. Being wrong about the behavioural assumption produces bias, which no quantity of data can fix. We therefore recommend looking for the presence of near-pure units in which a group's behaviour is nearly observed, thus allowing it to be held out and predicted in an out-of-sample procedure. Within our paper, this approach gave us measured error bars for every estimand where validation could be conducted, and flagged those where it could not (estimates for Indian and Other ethnic groups). Two features of the construction are conservative---the scored model sees less data than the one we report, and locality differences are left inside the measured error---while one runs the other way, since performance observed where the margins nearly pin a rate down is carried to streams where they do not. The absence of near-pure units for specific groups is itself highly informative---at a minimum, this should trigger closer inspection of the underlying data (for example, using simple scatter plots similar to those employed in this paper) to understand whether a particular point estimate is an extrapolation far beyond the support of the available data.

\section{Conclusion}
\label{sec:conclusion}

The July 2026 Johor landslide, decomposed against validated group-level estimates, uncovers critical information about contemporaneous electoral dynamics. There was a 28-point net reversion of Malay voters from PN to BN, robust to PN's stand-down in 23 seats. This reversion undid a green wave which had been steeply generational; young Malays carried it in 2022, their net reversal in 2026 was largest, and Malay politics flattened across age. Among Chinese voters who turned out, loyalty to PH largely persisted; demobilisation was the primary cause of PH's decline in Chinese support.

Our paper also contributes two durable artefacts. First, an open, reproducible pipeline for deriving bounded, validated estimates from saluran-level data---hitherto non-existent in empirical research on Malaysia. Second, a demonstration that the presence of near-pure units becomes a valuable resource allowing for out-of-sample validation when exploited. Beyond the specific estimates, these two artefacts are what we hope readers take away. As Malaysia approaches its 16th general election on the same roll and the same salurans, we hope our paper provides a rigorous baseline against which the next set of movements can be measured.

\section*{Data and code availability}
All inputs are public, from the Malaysian Election Corpus \cite{meco1}. Code for replication is available via \href{https://github.com/Thevesh/paper-saluran-jhr-se16}{GitHub}.

\section*{Funding}
This research did not receive any specific grant or funding.

\section*{Declaration of competing interests}

TT does not declare any competing interests. OKM is a member of the DAP---a component party of PH, one of the coalitions studied in this paper---and served as MP for Serdang (2013--2018) and subsequently Bangi (2018--2022). He left frontline politics in 2023 and is currently an Adjunct Professor at Taylor's University.

\section*{CRediT author statement}
Both authors contributed equally to the conceptualisation, design, and writing of this manuscript.

\newpage

\bibliography{references}

\end{document}